\documentclass[structabstract]{aa}  
\usepackage{graphicx}
\usepackage{txfonts}

\begin{document}

   \title{Synthetic photometry for carbon-rich giants}

   \subtitle{II. The effects of pulsation and circumstellar dust}

   \author{W.~Nowotny\inst{1}   \and
           B.~Aringer\inst{2}   \and
           S.~H\"ofner\inst{3}  \and
           M.T.~Lederer\inst{1}
           }

   \institute{University of Vienna, Department of Astronomy, 
              T\"urkenschanzstra{\ss}e 17, A-1180 Wien, Austria\\
              \email{walter.nowotny@univie.ac.at}
         \and
              INAF -- Padova Astronomical Observatory, 
              Vicolo dell'Osservatorio 5, 35122 Padova, Italy
         \and
              Department of Physics and Astronomy, 
              Division of Astronomy and Space Physics,
              Uppsala University,
              Box 516, SE-75120 Uppsala, Sweden\\
              %\email{}
             }

   \date{Received; accepted}

\titlerunning{Synthetic photometry for carbon-rich giants II.}
\authorrunning{W. Nowotny et al.}

% \abstract{}{}{}{}{} 
% 5 {} token are mandatory
 
\abstract
% context               heading (optional)
%                       {} leave it empty if necessary  
{Red giant stars approaching the end of the evolutionary phase of the Asymptotic Giant Branch (AGB) are, inter alia, characterised by (i) pulsations of the stellar interiors, and (ii) the development of dusty stellar winds. Therefore, such very evolved objects cannot be adequately described with hydrostatic dust-free model atmospheres.}
% aims                  heading (mandatory)
{By using self-consistent dynamic model atmospheres which simulate pulsation-enhanced dust-driven winds we studied in detail the influence of the above mentioned two effects on the spectral appearance of long period variables with carbon-rich atmospheric chemistry. While the pulsations lead to large-amplitude photometric variability, the dusty envelopes (resulting from the outflows which contain dust particles composed of amorphous carbon) cause pronounced circumstellar reddening.}
% methods               heading (mandatory)
{Based on one selected dynamical model which is representative of C-type Mira variables with intermediate mass loss rates, we calculated synthetic spectra and photometry for standard broad-band filters (Johnson-Cousins-Glass system) from the visual to the near-infrared. The synthetic photometry was subsequently compared with observational results.}
% results               heading (mandatory)
{Our modelling allows to investigate in detail the substantial effect of circumstellar dust on the resultant photometry. The pronounced absorption of amorphous carbon dust grains (increasing towards shorter wavelengths; $Q_{\rm abs}$\,/\,$a$\,$\propto$\,$\lambda^{-\beta}$ with $\beta$\,$\approx$\,1), leads to colour indices which are significantly redder than the corresponding ones based on hydrostatic dust-free models. Only if we account for this circumstellar reddening we get synthetic colours that are comparable to observations of evolved AGB stars. The photometric variations of the dynamical model were compared to observed lightcurves of the C-type Mira \object{RU Vir} which appears to be quite similar to the model (although the model is not a dedicated fit). We found good agreement concerning the principal behaviour of the $BVRIJHKL$ lightcurves and also quantitatively fitting details (e.g. magnitude ranges, the amplitude decrease from visual to NIR, absolute magnitudes). The analysed model is able to reproduce the variations of RU\,Vir and other Miras in ($J$\,--\,$H$) vs. ($H$\,--\,$K$) diagrams throughout the light cycle (ranges, loops). Contrasting the model photometry with observational data for a variety of galactic C-rich giants in such colour-colour diagrams proved that the chosen atmospheric model fits well into a sequence of objects with increasing mass loss rates, i.e., redder colour indices.}
% conclusions           heading (optional), leave it empty if necessary 
{The comparison of our synthetic photometry with observational results provides a further indication that the applied dynamic model atmospheres represent the outer layers of pulsating and mass-losing C-rich AGB stars reasonably well.}


   \maketitle

%####################################################################
\section{Introduction}
\label{s:intro}

In the first paper of this series, Aringer et al. (\cite{AGNML09}, from now on Paper\,I), we investigated synthetic spectra and photometry based on a grid of hydrostatic model atmospheres for C-rich red giant stars. A comparison with observational data demonstrated that these models are able to reproduce the emerging spectral energy distributions (SEDs) and the resulting fluxes in different photometric bands quite well, at least for objects with effective temperatures above $\approx$\,2800\,K. However (and not surprisingly), deviations are clearly visible for the very cool carbon stars as can be seen for example in Fig.\,14 of Paper\,I for the ($J$\,--\,$K$) colours. This is due to the fact that two significant phenomena, not covered by the basic assumptions of the COMARCS models, become important for the evolved stars on the Asymptotic Giant Branch (AGB), which is the evolutionary stage of the objects discussed here.

First, it is well known that the more evolved AGB stars show pronounced photometric variability caused by pulsations of the stellar interiors (e.g., Lattanzio \& Wood \cite{LattW04}). Already in 1596, David Fabricius reported the distinct light variation of a noticeable red star in the stellar constellation of the Whale which he named \textit{Stella Mirabilis}, i.e., the remarkable star. Considerable progress has been achieved since then in the field of stellar variability with $o$~Cet nowadays representing the prototype of a whole class of variable stars, the so-called long period variables (LPVs). They are characterised by variations with amplitudes of up to several magnitudes in the visual and on time scales of a few 10 to several 100 days. Historically, LPVs were subdivided into Mira variables, semiregular variables (SRV) and irregular variables (Lb) according to the regularity of the light change and the visual amplitude. As a consequence of the findings of the large variability surveys of the past (e.g., MACHO, OGLE) it is now considered being more appropriate to characterise LPVs by their pulsation mode, with Miras pulsating in the fundamental mode and the others in different overtone modes (e.g., Wood \cite{Wood00}, Ita et al. \cite{ITMNN04a}, Lebzelter \& Wood \cite{LebzW05}). While the amplitudes of Mira lightcurves reach several magnitudes in the visual (up to 10$^{\rm mag}$ as for example for $\chi$\,Cyg, Mattei \cite{Matte97}), they decrease further in the red ($<$\,3$^{\rm mag}$ in the \mbox{$I$-band,} e.g. Fig.\,5 in Hughes \& Wood \cite{HughW90}) and infrared (0.5--1$^{\rm mag}$, e.g. Fig.\,7 of Le\,Bertre \cite{LeBer92}). Bolometric light changes are of the order of 1$^{\rm mag}$ (e.g. Fig.\,5 in Whitelock et al. \cite{WhiMF00}).

Second, the stars on the upper part of the AGB lose mass via strong stellar winds with characteristic terminal velocities of 15\,km$\,$s$^{-1}$ and mass loss rates (MLRs) of typically 10$^{-6}$\,$M_{\odot}$\,yr$^{-1}$ (e.g., Olofsson \cite{Olofs04}, Fig.\,6 in Ramstedt et al. \cite{RamSO09}). Not only are AGB stars thereby contributing significantly to the enrichment of the surrounding ISM (e.g., Ferrarotti \& Gail \cite{FerrG06}), but the dusty outflows -- producing cool circumstellar envelopes -- have also a considerable effect on the spectral appearance. The interaction of the dust grains in the outflowing material with the radiation field of the star leads to distinctive spectral features in the mid- to far-infrared range (e.g., Posch et al. \cite{PKMDJ02}, Cami \cite{Cami02}, Molster \& Waters \cite{MolsW03}, Lebzelter et al. \cite{LPHWB06}, Cami \& Blommaert \cite{CamiB06}) as well as to a general circumstellar reddening due to the overall absorption from the visual to the infrared (IR) of common circumstellar dust species (e.g., Fig.\,1 of Wolfire \& Cassinelli \cite{WolfCC86}, Fig.\,1 of Groenewegen \cite{Groen06}, Andersen \cite{Ander07}). With increasing mass-loss rates $\dot M$ the SEDs as a whole become more and more redistributed from shorter to longer wavelengths (e.g., Fig.\,4 of Bedijn \cite{Bedij87}, Fig.\,1 of Sylvester et al. \cite{SKBJW99}, Fig.\,5.3 of Glass \cite{Glass99}, van Loon \cite{vanLo07}), the colour indices of such mass-losing red giants are considerably increased (e.g., van~der~Veen \& Habing \cite{VeenH88}, Le\,Bertre \& Winters \cite{LebeW98}) and the stars appear reddened.

Different dynamical modelling approaches have been used to model the photometric properties of evolved LPVs.

Based on the Berlin models (Fleischer et al. \cite{FleGS92}, Winters et al. \cite{WLJHS00}) for C-rich dust-driven winds, Winters et al. (\cite{WiFGS94}) presented theoretical light variations in the wavelength range of 0.5\,--25\,$\mu$m. They discussed the influence of dust formation processes on the shapes of the lightcurves, as for example the asymmetries during the pulsation cycle caused by newly emerging dust layers or the superposed long-term variations over a few cycles due to multiperiodicity effects in the dust-forming layers. The authors also confronted their results with observed JHKLM lightcurves for selected mass-losing C-type Miras showing qualitative agreement in the above mentioned points. Similar comparisons of synthetic lightcurves with observational photometric data were later published by Winters et al. (\cite{WiFLS97}) and for an M-type Mira by Jeong et al. (\cite{JeWLS03}).

By using the Australia-Heidelberg models (Hofmann et al. \cite{HofSW98}, Tej et al. \cite{TeLSW03}) for pulsating atmospheres of M-type Mira variables, Ireland et al. (\cite{IreSW04}) studied the effect of molecular absorption at different wavelengths in the NIR on the resulting light variations and compared synthetic JK lightcurves with observational data of O-rich Miras with quite moderate mass-loss rates. Very similar results were presented in Ireland et al. (\cite{IrSTW04}), also here the authors showed that their models were able to reproduce selected aspects of the observed photometric variabilities (not being the focus of either paper, though).

As described in detail in Nowotny et al. (\cite{NAHGW05}; Sect.\,3), both modelling approaches have their shortcomings when it comes to consistent and comprehensive modelling of the outer layers of evolved LPVs -- for example, the Berlin wind models are based on an insufficient description of the dust-free inner photospheric layers (no molecular opacities, grey radiative transfer) while the Australia-Heidelberg atmospheric models do not allow for dust formation and the resulting stellar wind. For the synthetic photometry presented here we applied the combined atmosphere and wind models of H\"ofner et al. (\cite{HoGAJ03}) which include these two points (non-grey molecular opacities and time-dependent treatment of dust-formation). In this way, they provide a detailed and consistent description of all the outer layers of an pulsating, mass-losing AGB star (cf. Nowotny et al. \cite{NowHA10}, below NHA10). In order to enable a direct comparison with observational data, we also chose a more subtle way to compute synthetic photometric magnitudes compared to the above mentioned studies, i.e., we used real filter systems instead of adopting simplified rectangular bandpasses, and we employed correspondingly the proper zeropoint magnitudes for an absolute calibration. Thereby, we carried on with the studies on synthetic photometry started by Windsteig et al. (\cite{WDHHK97}) and further pursued by Gautschy-Loidl et al. (Sect.\,4 of Gautschy-Loidl \cite{Gauts01}). The latter computed magnitudes and colour indices for a variety of different filter systems (among these also the one used here) based on C-rich model atmospheres. Most of these models were hydrostatic, but they also analysed a few proto-type non-grey dynamic model atmospheres. Significant improvements concerning the spectral synthesis (e.g., atomic lines) have been implemented since then (cf. Aringer et al. \cite{AGNML09}).

The aims for the work presented here were the following: 
(i) to point out the limited applicability of the static models from the grid presented in Paper\,I when dealing with very evolved red giant stars for which the assumptions of a hydrostatic configuration and no circumstellar dust are not necessarily valid anymore, 
(ii) to investigate the effects of pulsation and mass loss on the spectral and photometric appearance of AGB stars with the help of one selected state-of-the-art dynamical model atmosphere (Sect.\,\ref{s:modelling}), and
(iii) to test the quality of the applied dynamical models in representing the outer layers of mass-losing \mbox{C-rich} LPVs in a further and different way (compared to, e.g., the past spectroscopic studies; Gautschy-Loidl et al. \cite{GaHJH04}), namely via comparing of photometric modelling results with observational data (Sect.\,\ref{s:obsresults}).

For this purpose, we start with an investigation of the influence of different opacity sources on spectra and photometry in Sect.\,\ref{s:opacsources}, continue with the photometric temporal variations (lightcurves, changes in colour indices) in Sect.\,\ref{s:tempvar}, and present a detailed comparison with observed colour-colour diagrams in Sect.\,\ref{s:CMDCCD}.

%####################################################################
\section{Model atmospheres and spectral synthesis}
\label{s:modelling}

%********************************************************************
\subsection{Dynamic model atmospheres}
\label{s:DMAs}

\begin{table}
\begin{center}
\caption{Characteristics of the dynamic atmospheric model for a pulsating and mass-losing C-rich AGB star that was used for computing synthetic photometry.}
\begin{tabular}{llll}
\hline
\hline
&\multicolumn{2}{l}{Model:}& S \\
\hline
(i)&$L_\star$&[$L_{\odot}$]&10\,000\\
&$M_\star$&[$M_{\odot}$]&1.0\\
&$T_\star$&[K]&2600\\
&$[$Fe/H$]$&[dex]&0.0\\
&C/O ratio&\textit{by number $\qquad$}&1.4\\
\hline
(ii)&$R_\star$&[$R_{\odot}$]&493\\
&&[\textit{AU}]&2.29\\
&\multicolumn{2}{l}{log ($g_\star$ [cm\,s$^{-2}$])}&--\,0.94\\
\hline
(iii)&$P$&[d]&490\\
&$\Delta u_{\rm p}$&[km\,s$^{-1}$]&4\\
&$f_{\rm L}$&&2.0\\
&$\Delta m_{\rm bol}$&[mag]&0.86\\
\hline
(iv)&$\dot M$&[$M_{\odot}\,$yr$^{-1}$]&2.0\,-\,6.5\,$\times$\,10$^{-6}$\\
&$u$&[km\,s$^{-1}$]&13\,-\,15.5\\
&$\langle f_c$$\rangle$&&0.28\\
\hline
\end{tabular}
\label{t:dmaparameters}
\end{center}
\tablefoot{Listed are (i) parameters of the hydrostatic initial model, (ii) quantities derivable from these parameters, (iii) attributes of the inner boundary (piston) used to simulate the pulsating stellar interior as well as the resulting bolometric amplitude $\Delta m_{\rm bol}$, and (iv) properties of the resulting wind. The notation was adopted from previous papers (DMA3, NHA10): $P$, $\Delta u_{\rm p}$ -- period and velocity amplitude of the piston at the inner boundary; $f_L$ -- free parameter to adjust the luminosity amplitude at the inner boundary; $\dot M$, $u$ -- mass loss rate and outflow velocity at the outer boundary; $\langle f_c$$\rangle$ -- mean degree of condensation of carbon into dust at the outer boundary. The radial coordinates in Figs.\,\ref{f:massenschalenS} and \ref{f:structure} are plotted in units of the corresponding stellar radius $R_\star$ of the hydrostatic initial model, calculated from the luminosity $L_\star$ and temperature $T_\star$ via the relation $L_\star$\,=\,4$\pi$$R_\star^2$\,$\sigma$$T_\star^4$.}
\end{table}    

The synthetic photometry presented below is based on an atmospheric model which simulates the scenario of a pulsation-enhanced dust-driven wind as described in H\"ofner et al. (\cite{HoGAJ03}). The specific dynamical model used (\textit{model~S}) resembles a typical C-type Mira losing mass at an intermediate rate. In Table\,\ref{t:dmaparameters} the corresponding parameters are listed. This model proved to reproduce various observational results quite reasonably (e.g., NHA10), for details we refer to App.\,\ref{s:DMAdetails} (only available in the electronic edition).

Apart from the hydrostatic initial model (which is the starting point for the dynamic calculation), we studied atmospheric structures of model~S from three different and time-separated pulsation cycles -- namely $\phi_{\rm bol}$\,=\,[{\it 0,1}], [{\it 7,8}], [{\it 14,15}] -- with 23 individual dynamic phases for each cycle (cf. Fig.\,8 in Nowotny et al. \cite{NoLHH05}). This allowed to study the variability throughout a light cycle as well as any cycle-to-cycle variations. 

Following the convention of previous papers (e.g., NHA10), the modelling results will be marked throughout this work by bolometric phases $\phi_{\rm bol}$ within the light cycle in luminosity (cf. Fig.\,\ref{f:structure}a) with numbers written in \textit{italics}, in contrast to the visual phases $\phi_{\rm v}$, which are usually used to designate observational results (with $\phi_{\rm v}$=0 corresponding to phases of maximum light in the visual). The difficulties emerging when the two kinds of phase informations are to be related were discussed in detail in App.\,A of NHA10.

%********************************************************************
\subsection{Synthetic spectra}
\label{s:synthspec}

The outcome of the dynamic atmospheric modelling are snapshots of the radial atmospheric structure during the temporal evolution as shown in Fig.\,\ref{f:structure}. These served as input for the subsequent detailed a posteriori radiative transfer calculations. 

Based on the given $T$-$p$-structures, opacities were then calculated with the {\tt COMA} code which was already described in detail in Paper\,I. Additional information concerning this opacity generation code and the various improvements in the past years can be found in Aringer (\cite{Aring00}), Gautschy-Loidl (\cite{Gauts01}), Gorfer (\cite{Gorfe05}), Cristallo et al. (\cite{CrSLA07}), Lederer \& Aringer (\cite{LedeA09}), Aringer et al. (\cite{AGNML09}), and Nowotny et al. (\cite{NowHA10}). In accordance with the dynamical model (cf. Table\,\ref{t:dmaparameters}), we also assumed -- except for carbon (see below) -- solar-like elemental abundances for the spectral synthesis. Aringer (\cite{Aring05}) pointed out, that it is important to be consistent in this point. We adopted the values from Anders \& Grevesse (\cite{AndeG89}), except for C, N and O where we took the data from Grevesse \& Sauval~(\cite{GrevS94}). This agrees with our previous work (e.g., Aringer et al.~\cite{AHWHJ99}, Paper\,I) and results in a $Z_{\odot}$ of $\approx$\,0.02. Starting from the adopted solar abundance pattern, the carbon abundance was subsequently increased according to the C/O ratio of the model (without changing the rest of the composition). Abundances of various important neutral and ionised species (atomic, molecular) were then computed for all layers of the model atmosphere assuming chemical equilibrium for molecule formation and equilibrium for ionisation (see Lederer \& Aringer \cite{LedeA09} for a detailed discussion and an extensive list of references). In addition, the depletion of the element C from the gas phase due to consumption by formation of amC dust grains was accounted for. In the next step, the resulting abundances -- i.e. partial pressures of the neutral atoms, ions, and molecules as a function of atmospheric depth (cf. Fig.\,2 of Loidl et al. \cite{LoHJA99} or Fig.\,5 of NHA10) -- can be used to calculate absorption coefficients for every point of a chosen grid in wavelength~$\lambda$ and under the assumption of LTE. All necessary opacity sources were considered: \\
\hspace*{0.8cm} - \hspace*{0.02cm} continuous, \\
\hspace*{0.8cm} - \hspace*{0.02cm} lines (=\,atoms\,+\,molecules), and \\
\hspace*{0.8cm} - \hspace*{0.02cm} dust.

\begin{figure}
\resizebox{\hsize}{!}{\includegraphics[clip]{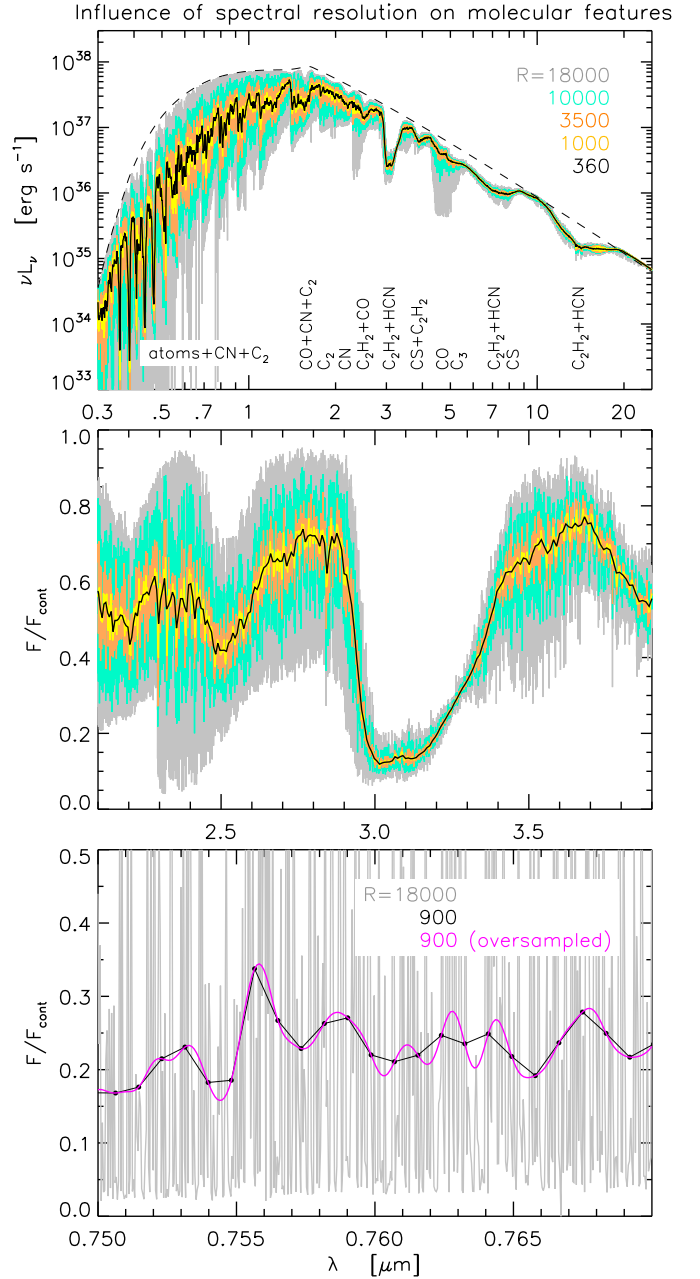}}
\caption{Synthetic spectrum based on the hydrostatic and dust-free initial model ($R$\,=\,18\,000, grey) calculated as described in Sect.\,\ref{s:synthspec}, and some spectra of lower resolutions (as given in the legend) to illustrate the effect of the rebinning on the appearance of the molecular features. In the top panel, the whole spectral range is shown, while the middle panel show the details of a smaller sections. The dashed line marks a calculation where all line contributions were neglected (theoretical continuum level). The bottom panel shows the difference for a spectrum of lower resolution if oversampled by a factor of 10 (magenta) or not (black).}
\label{f:specreseffects}
\end{figure}

For the continuous absorption all the sources listed in Lederer \& Aringer (\cite{LedeA09}) were taken into account. 
We also followed their approach concerning the atomic lines with line data taken from the VALD database (Kupka et al. \cite{Kupka00}). However, for the sake of computing time only Doppler profiles were adopted for the description of the line shapes (see Sect.\,2.4 in NHA10 for details) instead of full Voigt profiles. Test calculations have shown that this simplification has only a marginal effect on the synthetic photometry (e.g., differences in the second decimale place for the $V$-magnitude) while the computing time is smaller by a factor of two. For the microturbulence velocity we assumed $\xi$\,=\,2.5\,km\,s$^{-1}$, which is in line with our previous work (Aringer et al. \cite{AriJL97}, \cite{AHWHJ99}) and also with observational results for AGB stars (e.g., Smith \& Lambert \cite{SmitL90}, Lebzelter et al. \cite{LLCHS08}), as well as it is consistent with the values that were used for preparing the gas opacity tables for the hydrodynamic modelling.
For the molecular contributions we did not use line lists\footnote{The set of molecular species included in the current version of the COMA code and the corresponding references of the line lists can be found in Lederer \& Aringer (\cite{LedeA09}), with recent updates described in Aringer et al. (\cite{AGNML09}).} as in Paper\,I, but used the same opacity tables\footnote{Mainly from the SCAN data base (J{\o}rgensen \cite{Jorge97}; see also \textrm{http://www.astro.ku.dk/$\sim$uffegj/}), a detailed description of which can be found in J{\o}rgensen et al. (\cite{JorgJ93}, \cite{JoJSA01}). For a discussion of the original OS data and revisions we refer to Gautschy-Loidl (\cite{Gauts01}, Sects.\,2.1+2.2).} as Gautschy-Loidl et al. (\cite{GaHJH04}), accounting for the following C-bearing molecules: CO (Goorvitch \& Chackerian \cite{GoorC94}), CN (J{\o}rgensen \& Larsson \cite{JorgL90}), C$_2$ (Querci et al. \cite{QueQT74}), CH (J{\o}rgensen et al. \cite{JoLIY96}), HCN (J{\o}rgensen et al. \cite{JAGLS85}), C$_2$H$_2$ (J{\o}rgensen \cite{Jorge97}), and C$_3$ (J{\o}rgensen et al. \cite{JorAS89}). An overview of the individual spectral features of the different molecular species can be found in Fig.\,4 of Paper\,I or in Gautschy-Loidl et al. (\cite{GaHJH04}, Fig.\,1). The significant contributions in the diverse wavelength regions are also identified in the upper panel of Fig.\,\ref{f:specreseffects}.
Finally, the absorption due to dust grains of amorphous carbon was -- in consistency with the preceding modelling of the atmospheric structures -- calculated using the data given in Rouleau \& Martin (\cite{RoulM91}; set AC, cf. bottom panel of Fig.\,\ref{f:specdusteffects}) and assuming the small particle limit of the Mie theory to be valid.\footnote{The influence of using different dust data (opacity data sets, grain densities, etc.) when modelling C-type LPVs with dusty outflows was already discussed in Andersen et al. \cite{AndHG03}). There the effect on the resulting winds (mass loss rates, wind velocities, condensation degrees) as well as on the resulting colour indices was demonstrated.}

Summing up the opacity contributions of all sources, we computed combined opacity tables $\kappa_{\nu}$($\lambda$,r) in the wavelength range of 0.2\,--\,25\,$\mu$m with a spectral resolution of $R$\,=\,18\,000 (as determined by the above listed molecular opacity tables). On the basis of these the radiative transfer (RT) was solved for every wavelength point $\lambda_{i}$ in spherical geometry (e.g., Yorke \cite{Yorke80}, Nordlund \cite{Nordl84}) with the help of our dedicated RT code (Balluch \cite{Ballu86}, Windsteig \cite{Winds95}), which was originally developed for the studies presented in Windsteig et al. (\cite{WDHHK97}). Note that no velocity effects were taken into account in the a posteriori RT (Doppler shifts), in contrast to the recent work published in Nowotny et al. (\cite{NowHA10}). 

The resulting spectra served then as input for calculating photometric magnitudes as described in the following Sect.\,\ref{s:synthphot}. With the spectral resolution of $R$\,=\,18\,000 there are statistically meaningful numbers of points (a few $\times$\,10$^3$) in the wavelength ranges of the chosen broad-band filters. 
However, the resolution of the synthetic spectra is much lower than what would be needed for an appropriate description of individual lines (the typical line widths require a sampling with $R$ larger than $\approx$100.000). Thus, our approach works similar to the opacity sampling (OS) method.\footnote{In the OS technique (e.g., Ekberg et al. \cite{EkbEG86}, Alexander et al. \cite{AleAJ89}, J{\o}rgensen \cite{Jorge92}), the mono-chromatic molecular absorption is computed at a limited number of wavelength points, distributed over the spectral range considered (more or less randomly, but not in such numbers that the spectrum is resolved). Opacities at the selected wavelength points are directly calculated from line lists and subsequently used to solve the radiative transfer. Due to the statistical nature of the OS approximation, it is necessary to average the OS spectrum over a number of frequency points in order to produce realistic spectra and reproduce observed features (e.g., Gautschy-Loidl \cite{Gauts01}, Aringer et al. \cite{AGNML09}). As a consequence, the OS technique is in general only applied for spectroscopic studies with low spectral resolutions.} Like OS spectra, our computed spectra have a statistical nature and need to be averaged over a certain number of wavelength points ($\approx$\,20\,--100) for further spectroscopic investigations (e.g., comparison with observational data). The influence of such a decrease in spectral resolution on absorption features is demonstrated in Fig.\,\ref{f:specreseffects}, its severeness changes with wavelength (compare, e.g., in the middle panel $\lambda$\,=\,2.4\,$\mu$m and 3.3\,$\mu$m). In this case, realistic representations of observed spectra have resolutions below $R$\,=\,1000 (yellow, black). For the plots containing spectra below (Figs.\,\ref{f:specdusteffects}+\ref{f:specvariationDMA}), the original spectra were averaged over 50 points leading to a reduced spectral resolution of $R$\,=\,360. In addition, the sampling rate of the output wavelength grid should be higher (roughly by a factor of $\approx$\,5\,--10) than the nominal resolution of the spectrum with decreased resolution, in order to obtain a smooth spectrum (similar to spectrographs; cf. sampling theorem). In the bottom panel of Fig.\,\ref{f:specreseffects}, the effect of this oversampling on the appearance is illustrated by a comparison of a low-resolution spectrum with $R_{\rm nominal}$\,=\,$R_{\rm sampling}$\,=\,900 (black) with the corresponding spectrum having $R_{\rm nominal}$\,=\,900 and $R_{\rm sampling}$\,=\,9000 (magenta).

%********************************************************************
\subsection{Synthetic photometry}
\label{s:synthphot}

To compute magnitudes on the basis of the synthetic spectra we needed filter transmissions as a function of wavelength to convolve with and associated calibration standards (i.e., photometric zeropoints). From the large number of photometric systems in use (e.g., Bessell \cite{Besse05}) we chose standard broad-band filters to facilitate the comparison of modelling results with observational findings. Thus, we utilised as in Paper\,I the response curves $R(\lambda)$\,$\in$\,$[0,1]$ for the following filters: Johnson-Cousins UBV and Cousins RI as given in Bessell (\cite{Besse90}), and JHKLL$^\prime$M from the homogenized Johnson-Glass system described in Bessell \& Brett (\cite{BB88}). The data were retrieved in electronic format from the {\tt Asiago Database on Photometric Systems} (see Moro \& Munari \cite{MoroM00}). Bessell et al. (\cite{BesCP98}) quote the corresponding absolute fluxes $f_{\lambda}^0$ and zeropoint magnitudes $m^0$ for the standard Johnson-Cousins-Glass UBVRIJHKLL$^\prime$M system. Those were derived from models of Vega (Castelli \& Kurucz \cite{CastK94}) and Sirius (as for example in Cohen et al. \cite{CoWBD92}) and observed colours of these stars.

The magnitude $m_X$ for a particular filter $X$ out of the above mentioned set is calculated by means of (see footnote of Table~A2 in Bessell et al. \cite{BesCP98})
\begin{equation}\label{eq:magnitude}
   m_X=-2.5\log f_{\lambda,X}-21.1-m_X^0\mathrm{,}
\end{equation}
whereby
\begin{equation}\label{eq:meanflux}
   f_{\lambda,X}\equiv\frac{\int_{\lambda_{X,l}}^{\lambda_{X,u}}f_\lambda(\lambda) R_X(\lambda) 
   d\lambda}{\int_{\lambda_{X,l}}^{\lambda_{X,u}} R_X(\lambda) d\lambda}
\end{equation}
and $f_\lambda$ is the energy flux per unit area and wavelength interval. The lower and upper cut-off wavelength of the filter passband are represented by the quantities $\lambda_{X,l}$ and $\lambda_{X,u}$, respectively. Setting $f_{\lambda,X}$\,=\,$f_{\lambda,X}^0$ in Eq.~\ref{eq:magnitude} leads to $m_X$\,=\,$0$.

The absolute fluxes $f_{\lambda}^0$ refer to values measured at the Earth. Our synthetic spectra specify the flux at the stellar surface $F_\lambda^\star (\lambda)$. To deduce absolute magnitudes we have to ``shift'' the star to a distance of $d$\,=\,$10\,\mathrm{pc}$. Ignoring the effects of the interstellar extinction, we get
\begin{displaymath}
   m-M=5\log \,d \, \mathrm{[pc]}-5=0
\end{displaymath}
and, thus, the apparent magnitude $m$ equals the absolute magnitude $M$. The stellar flux $f_\lambda^\star$ at a distance $d$ emerges from the identity
\begin{displaymath}
   L_\lambda=4\pi R^2 F_\lambda^\star = 4\pi d^2 f_\lambda^\star
\end{displaymath}
with $R$ depicting the stellar radius.

With the prescription given in Eq.~\ref{eq:meanflux} we calculate $f_{\lambda,X}^\star$ from $f_\lambda^\star$. We choose to infer the magnitude by setting $f_{\lambda,X}^\star$ into relation to the zeropoint flux $f_{\lambda,X}^0$ rather than inserting $f_{\lambda,X}^\star$ directly into Eq.~\ref{eq:magnitude}. That way the additive constants cancel out, i.e., 
\begin{displaymath}
   m_X=-2.5\log f_{\lambda,X}^\star-21.1-m_X^0-\underbrace{(-2.5\log f_{\lambda,X}^0-21.1-m_X^0)}_{=0}\textrm{,}
\end{displaymath}
and we arrive at
\begin{equation}\label{eq:synthmagnitude}
   M_X=m_X=-2.5\log \,(f_{\lambda,X}^\star/f_{\lambda,X}^0)\textrm{.}
\end{equation}

The desired synthetic absolute magnitudes (e.g., Fig.\,\ref{f:lightcurvesDMA}) could then be computed with the help of Eq.\,\ref{eq:synthmagnitude}, where absolute fluxes $f_{\lambda,X}^0$ for a fictitious A0 reference star were taken from the literature (Table~A2 in Bessell et al. \cite{BesCP98})\footnote{Note that the magnitude zeropoints zp(f$_{\lambda}$) and zp(f$_{\nu}$) in this table are erroneously reverse labelled as stated by Bessell (\cite{Besse05}).} and the filter fluxes of our model $f_{\lambda,X}^\star$ were determined by convolving the synthetic spectra with the corresponding filter transmission curves $R_X(\lambda)$ according to Eq.\,\ref{eq:meanflux}.

For test purposes we computed synthetic photometry in the way sketched above for the grid of hydrostatic COMARCS models presented in Paper\,I and compared the values with the photometry given there %(CDS) 
which was calculated following the more sophisticated approach described in Girardi et al. (\cite{GBBCG02}). Averaged over 738 models we found mean differences of $\approx$\,0$\fm$024\,--\,0$\fm$077 for the filters VRIJHKL. For the M filter we found a larger difference of 0$\fm$13, while no direct comparison could be carried out for the B filter due to the missing wavelength coverage of the spectra in Paper\,I. Note that these differences are largely caused by a systematic shift in the zeropoints, while the standard deviations of the magnitude difference distributions for most filters are below $\approx$\,0$\fm$009.

A note concerning the scaling of C$_2$ opacities and the resulting shift in ($H$\,--\,$K$): As described by Loidl et al. (\cite{LoiLJ01}), the $gf$ values of the C$_2$ lines in the list of Querci et al. (\cite{QueQT74}) were scaled down beyond 1.15\,$\mu$m for the preparation of their OS tables. According to Loidl et al. this should provide a more realistic description of the C$_2$ absorption compared to the unscaled Querci list. In Paper\,I we found that the applied scaling leads to moderate changes of the spectra in the region of \mbox{1.3\,--\,2.1\,$\mu$m}, i.e., mainly in the H-band. From a comparison of observations of C stars with synthetic photometry based on hydrostatic models and computed with scaled C$_2$ data (see Fig.\,15 in Paper\,I) it was suspected that this correction leads to worse \mbox{($H$\,--\,$K$)} colour indices. As a rule of thumb, it was recommended in Paper\,I to add 0$\fm$1 to the synthetic ($H$\,--\,$K$) colours of the hydrostatic grid to compensate for this alleged deficient scaling (equal to using the original line list). This was done for the hydrostatic models in Figs.\,\ref{f:JHKObsMAERCSE} and \ref{f:JHKObsDMA}. For the spectral synthesis based on dynamic models presented here we used the same OS tables as Loidl et al. (\cite{LoiLJ01}). However, no investigations on the differences in \mbox{($H$\,--\,$K$)} colours were made for the dynamic models as the effects of pulsation and circumstellar dust (cf. Fig.\,\ref{f:colorcurvesDMA}) are significantly larger than arguable changes due to uncertain molecular line lists.

To allow for a direct comparison with the modelling results, all observational data (Sect.\,\ref{s:obsresults}) were, if necessary, transformed to the Johnson-Cousins-Glass or Bessell photometric system which was used for the calculation of synthetic magnitudes as just described and should serve as the standard system throughout this work.

%********************************************************************
\section{Observational results for comparison}
\label{s:obsresults}

In order to test the ability of our models to reproduce the photometric appearance of real C-rich LPVs we used observational data from the literature. A valuable source in this respect was the compilation of near-IR data for a large sample of galactic carbon-rich variable AGB stars presented by Whitelock et al. (\cite{WhFMG06}). They distinguished between Miras with clearly periodic lightcurves and peak-to-peak amplitudes in $K$ larger than 0.4$^{\rm mag}$ (i.e., their variability type 1n) and other LPVs with variations that are not periodic or have amplitudes less than 0.4$^{\rm mag}$ (i.e., their class 2n for non-Miras). On top of the photometric time-series in JHKL, they also derived for a subsample of C-rich Miras other quantities which are useful for a detailed comparison with modelling results: estimates for the distance of the targets, the interstellar reddening in the corresponding direction, bolometric magnitudes and corrections, or estimates for the mass-loss rates.

We chose one specific object from the sample of Whitelock et al. (\cite{WhFMG06}) -- namely the \mbox{C-type} Mira RU Vir -- for a more detailed comparison as the known properties of this star (listed in Table\,\ref{t:RUVirparameters}) are roughly comparable to the model parameters (Table\,\ref{t:dmaparameters}) and it showed a photometric behaviour quite similar to our model~S. However, model~S was not intended to be a specific fit for RU\,Vir. The difficulties arising when we aim at relating dynamic model atmospheres to certain observable targets were discussed in Nowotny et al. (\cite{NoLHH05}). Crucial for the choice of RU\,Vir was, in addition, the fact that it is one of the few LPVs for which also time-series photometry in the visual and red is avaible (although obtained at a different interval in JD than the Whitelock observations). The respective BVRI lightcurves were published in Eggen (\cite{Eggen75a}) and analysed by Eggen (\cite{Eggen75b}). Furthermore, the quite low interstellar extinction in the direction of RU\,Vir (Table\,\ref{t:RUVirparameters}) is convenient.

\begin{table}
\begin{center}
\caption{Observed and derived properties of the C-type Mira RU\,Vir collected from the literature.}
\begin{tabular}{llll}
\hline
\hline
&&&Refs.\\
\hline
$L$ & [$L_{\odot}$] & 6888 & 1 \\
$m_{\rm bol}$ & [mag] & 4.94 & 1 \\
$K_{\rm 0}$ & [mag] & 1.79 & 1\\
distance & [pc] & 910 & 1 \\
($m$\,--\,$M$) & [mag] & 9.80 & 1 \\
$A_V$ & [mag] & 0.08 & 1 \\
spectral type & & C\,8,1e (R3ep) & 2 \\
\hline
$P_{K}$ & [d] & 444 & 1 \\
$P$ & [d] & 433.2 & 2 \\
$\Delta V$ & [mag] & 5.2 & 2 \\
           &       & 6.7 & 3 \\
$\Delta K$ & [mag] & 0.98 & 1 \\
\hline
$\dot M$ & [$M_{\odot}$\,yr$^{-1}$] & 2.0\,-\,8.3\,$\times$\,10$^{-6}$ & 1,4,5,6 \\
$v_{\rm exp}$ & [km\,s$^{-1}$] & 16.9\,-\,18.4 & 4,5,6 \\
\hline
\end{tabular}
\label{t:RUVirparameters}
\end{center}
\tablebib{(1)~Whitelock et al. (\cite{WhFMG06}); (2)~GCVS, Samus et al. (\cite{SamuD09}); (3)~AAVSO data ($V_{\rm max}$\,--\,$V_{\rm min}$); (4)~Bergeat \& Chevalier (\cite{BergC05}); (5)~Loup et al. (\cite{LoFOP93}); (6)~Knapp \& Morris (\cite{KnapM85}).}
\end{table}      

From Whitelock et al. (\cite{WhFMG06}) we adopted for RU\,Vir a series of 46 data points in JHKL over a time span of $\approx$\,17 years (i.e., $\simeq$\,14 pulsation periods). The observed magnitudes were not transformed as the differences between the SAAO photometric system and the Bessell filters we used for the modelling (Sect.\,\ref{s:synthphot}) are almost negligible according to Bessell \& Brett (\cite{BB88}, Table\,I).\footnote{Especially when compared with the dynamic effects we focus on in this work. While, for example, the difference between ($J$\,--\,$K$)$_{\rm init}$ and \mbox{($J$\,--\,$K$)$_{\rm dyn}$} amounts to 2\,-\,3$^{\rm mag}$ (cf. Fig\,\ref{f:colorcurvesDMA}), the transformation would be ($J$\,--\,$K$)$_{\rm BB}$ = 1.0\,($J$\,--\,$K$)$_{\rm SAAO}$ -- 0.005.}

Eggen (\cite{Eggen75a}) lists 27 measurements in BV as well as 23 in RI over an interval of $\approx$\,2.5 years (i.e., about 2 periods). The BV magnitudes are given in the Cousins system (Eggen \cite{Eggen75c}), which is essentially identical to BV in the Johnson system (Bessell \& Weis \cite{BessW87}), and could, therefore, directly be used here. For a comparison of the RI magnitudes with our synthetic photometry we needed to transform these from the (Eggen-)Kron system to the Cousins system with the help of the following conversions which are based on the data of Bessell \& Weis (\cite{BessW87}) and were kindly provided by M.S.~Bessell (priv. comm.):
\begin{eqnarray}
  \begin{array}{lll}
   \vspace{0.2cm}
   I_C & = & I_K - 0.227 + 0.0367 \, (R-I)_K \\
   R_C & = & R_K - 0.113 - 0.0609 \, (R-I)_K + 0.33 \, (R-I)_K^2 - \nonumber \\ 
       &   & - \, 0.085 \, (R-I)_K^3
  \end{array}
\end{eqnarray}

\begin{figure}
\resizebox{\hsize}{!}{\includegraphics[clip]{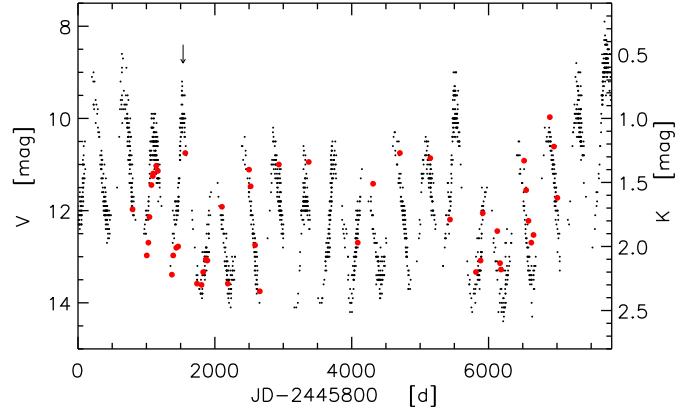}}
\caption{Light variations of the C-type Mira RU\,Vir in the visual (AAVSO data; black dots) and in the K-band (adopted from Whitelock et al. \cite{WhFMG06}; red points). The arrow marks JD\,2447336 which was determined to be a maximum phase ($\phi_{\rm v}$=0) from the AAVSO data and then adopted to estimate phases for the NIR data (see text). On top of the regular variation some long-term trend can be recognised (cf. Fig.\,2b in Mattei \cite{Matte97}).}
\label{f:lightcurves}
\end{figure}

Our aim was to combine the available photometric observations of RU\,Vir -- coming from different light cycles sampled irregularly and insufficiently (Fig.\,\ref{f:lightcurves}) -- into one composite light cycle with $\phi_{\rm v}$\,$\in$\,[0,1], leading to well-sampled lightcurves as a function of phase (Fig.\,\ref{f:lightcurvesRUVir}). However, this kind of phase information is not given by Whitelock et al. (\cite{WhFMG06}) for the NIR data and the values listed by Eggen (\cite{Eggen75a}) appear doubtful as the light maximum clearly deviates from $\phi_{\rm v}$=0 (cf. Fig.\,2 in Eggen \cite{Eggen75b}). Therefore, we used the  photometric data in the visual available for RU\,Vir in the AAVSO database\footnote{http://www.aavso.org/data/lcg} (Fig.\,\ref{f:lightcurves} or Fig.\,2b in Mattei \cite{Matte97}) to derive visual phases ourselves. With the help of {\tt Period04}, a software to analyse photometric time series (Lenz \& Breger \cite{LenzB05}), we made a Fourier-fit to the AAVSO V-band lightcurve in the range where the observations of Eggen (\cite{Eggen75a}) and Whitelock et al. (\cite{WhFMG06}) were carried out (the derived period of 434$^{\rm d}$ is quite similar to the values in the literature, cf. Table\,\ref{t:RUVirparameters}). Thereby, we could determine the points in time of maximum phase for both time series to JD\,2441262 and 2447336, respectively. This allowed to compute phases $\phi_{\rm v}$ for each measurement and plot the latter co-phased as shown in Fig.\,\ref{f:lightcurvesRUVir}. Subsequently, we computed sinusodial fits (one Fourier component) to the resulting combined lightcurves in all filters, again using {\tt Period04}. The fits are shown in Fig.\,\ref{f:lightcurvesRUVir}, too.

Apart from the time-series photometry we also used data from a few other sources. The observational data adopted from Bergeat et al. (\cite{BerKR01}; Table\,4) in Fig.\,\ref{f:JHKObsMAERCSE} were compiled by those authors from the literature. According to Knapik et al. (\cite{KnapB97}) the measurements were obtained in the "Arizona system or close to it" (i.e., Johnson), and can be directly compared (Bessell \& Brett \cite{BB88}, Table\,I). 

In addition, a representative sub-sample of the extensive grid of hydrostatic model atmospheres of Paper\,I (covering the whole range in colours) was selected and the corresponding photometry utilised for Fig.\,\ref{f:JHKObsMAERCSE}.

\begin{figure}
\resizebox{\hsize}{!}{\includegraphics[clip]{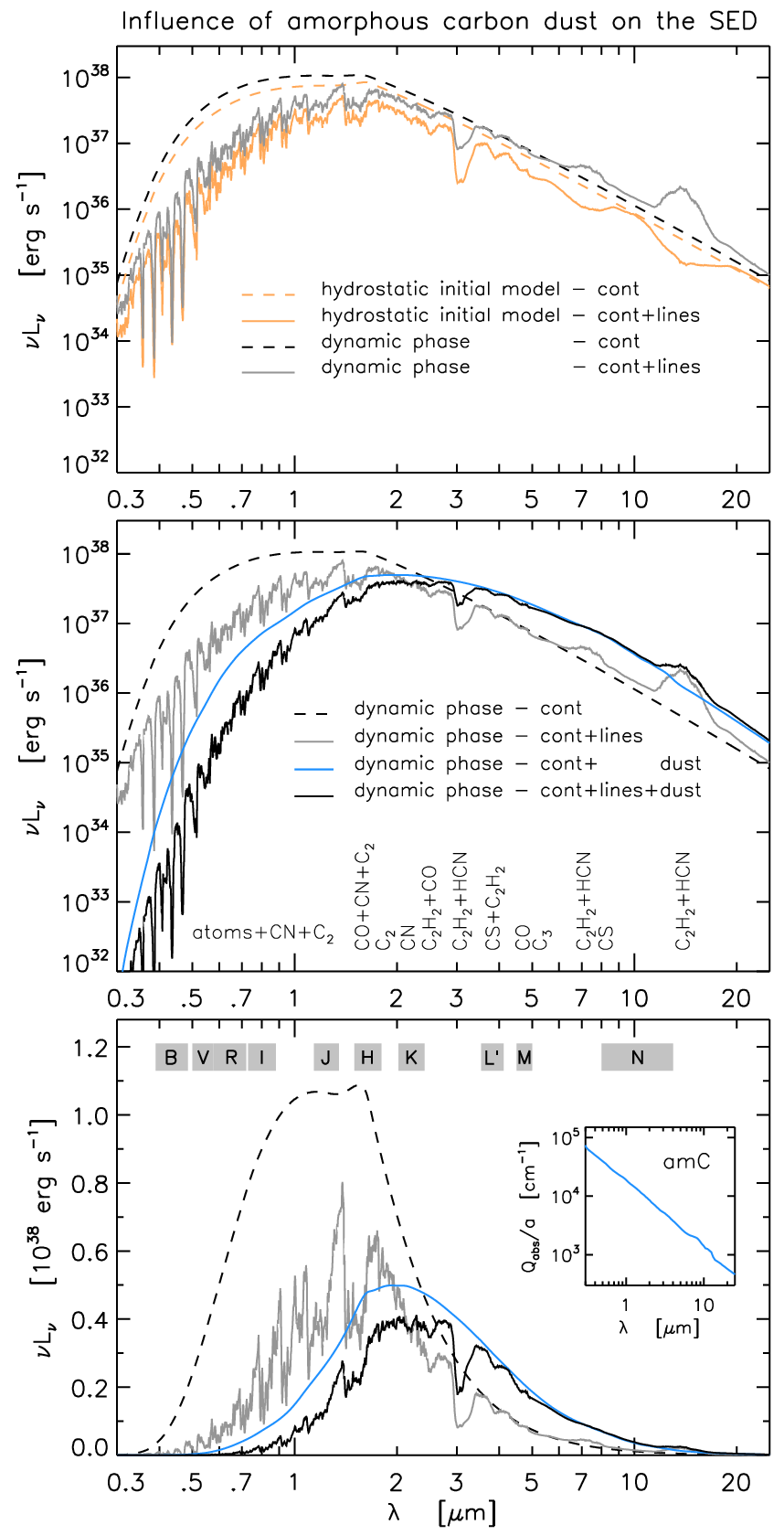}}
\caption{Illustration of the influence of different opacity sources (for details see text) -- especially the grains of amorphous carbon dust -- on the resulting spectra demonstrated with the help of the dust-free initial model and one phase of the dynamic calculation ($\phi_{\rm bol}$\,=\,\textit{0.0}) of model~S. Different sources were included in the a posteriori RT as denoted in the legend. In the middle panel the molecular species responsible for the major features are indicated. In the bottom panel (showing the same as the middle panel but scaled linearly) the wavelength ranges of the used broad-band filters ($\approx$\,FWHM of the responses given in Bessell \cite{Besse90} and Bessell \& Brett \cite{BB88}) are marked for orientation purposes, while the insert shows the absorption data for amorphous carbon dust from Rouleau \& Martin (\cite{RoulM91}) as applied for the computations.}
\label{f:specdusteffects}
\end{figure}

%####################################################################
\section{Modelling results}
\label{s:results}

%********************************************************************
\subsection{The extended atmosphere and the role of dust}
\label{s:opacsources}

\begin{figure}
\resizebox{\hsize}{!}{\includegraphics[clip]{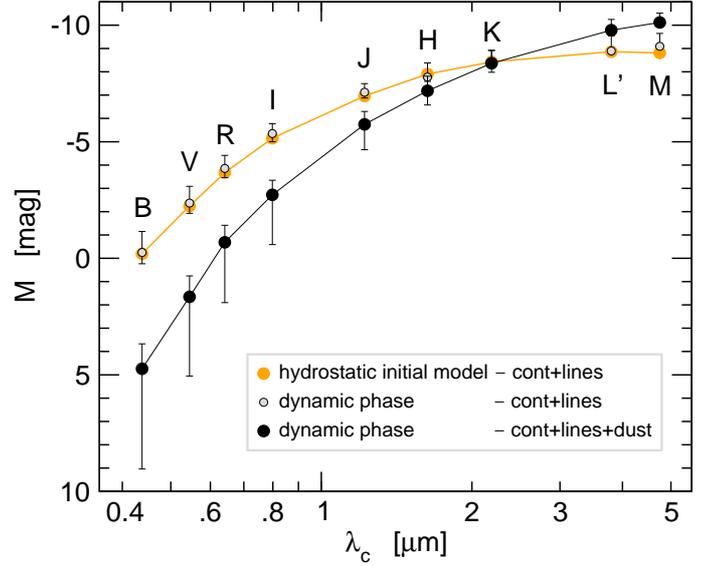}}
\caption{The influence of different opacity sources on the broad-band photometric magnitudes demonstrated with the help of the initial model and one selected dynamical phase ($\phi_{\rm bol}$\,=\,\textit{0.31}) of model~S. In addition, the maximum variations throughout the pulsation cycle are marked with bars. Same colour code as in Fig.\,\ref{f:specdusteffects}.}
\label{f:photdusteffects}
\end{figure}

\begin{figure}
\resizebox{\hsize}{!}{\includegraphics[clip]{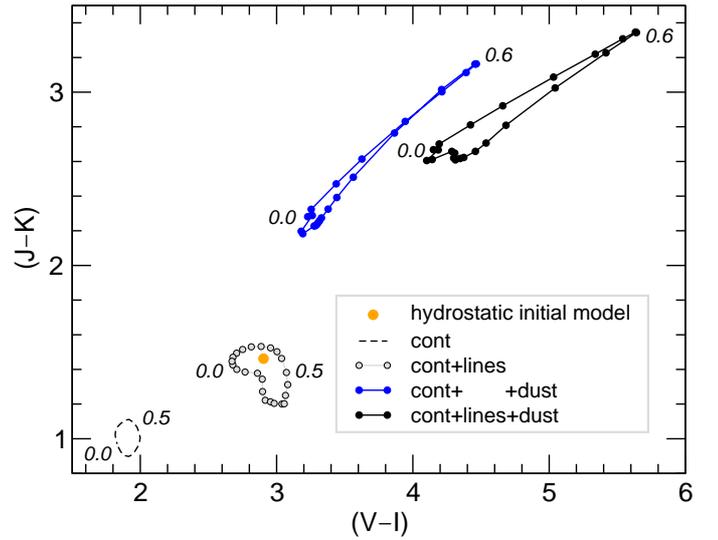}}
\caption{The influence of different opacity sources on the location in a colour-colour diagram. Shown are the hydrostatic initial model and several dynamical phases throughout a pulsation cycle of model~S with opacity sources accounted for as listed in the legend. Same colour code as in Fig.\,\ref{f:specdusteffects}. Selected phases $\phi_{\rm bol}$ are denoted for orientation.}
\label{f:photdusteffects1}
\end{figure}

The differences between the relatively compact hydrostatic and dust-free initial model and the developed model with dust particles occuring in the stellar wind (Fig.\,\ref{f:structure}) shall be discussed in the following, as well as the effects of the different opacity sources on spectra and photometry. In Fig.\,\ref{f:specdusteffects} we compare synthetic low-resolution spectra (Sect.\,\ref{s:synthspec}) based on the initial model and one phase of the dynamic model for which the different sources of opacity were taken into account in the a posteriori radiative transfer or not.\footnote{The latter is not consistent with the dynamical modelling (no conservation of energy, i.e. not the same total flux) but useful for illustration purposes.} Our spectra cover the wavelength range where the bulk of the flux is emitted by the star, with the SED peaking around 1--3\,$\mu$m. Across the whole range significant contributions of (atomic and) molecular features are found, which is characteristic of late-type giants. On top of the absorption\footnote{Concerning the molecular features seen in emission around $\approx$\,7$\mu$m and $\approx$\,14$\mu$m we refer to the discussions in Gautschy-Loidl (\cite{Gauts01}; Sect.\,3.3) and Gautschy-Loidl et al. (\cite{GaHJH04}; Sect.\,6), as this effect is not relevant for the photometric results in this work.} due to molecules, there is also the severe effect due to amorphous carbon dust in the outflow ($>$\,2\,$R_\star$; cf. Fig.\,\ref{f:structure}f) clearly visible. The absorption features in the visual and red (mainly CN, C$_2$) appear attenuated, this can become more pronounced for other phases of the pulsation cycle as can be seen in Fig.\,\ref{f:specvariationDMA}. Even more noteworthy is the overall shift of the SED towards longer wavelengths. As the circumstellar amC dust grains efficiently absorb radiation of the star in the visual ($\propto$\,$Q^{\rm abs}_{\nu}$$J_{\nu}$) and re-emit it in the IR ($\propto$\,$Q^{\rm abs}_{\nu}$$B_{\nu}(T_{\rm d})$), the flux is considerably redistributed. In this way, the dust is not only the trigger for the stellar wind, but has also a pronounced effect on the spectral appearance.

This noticeable circumstellar reddening can also be recognised in Fig.\,\ref{f:photdusteffects}. There the photometric magnitudes for the chosen standard filters (Sect.\,\ref{s:synthphot}) based on the spectra of Fig.\,\ref{f:specdusteffects} are plotted. It is interesting to see that only small differences can be found between the SED based on the hydrostatic model atmosphere and the SED for the dynamic phase for which only continuous and line contributions (atoms, molecules) were taken into account while dust absorption was neglected. On the other hand, a significant dimming of a few magnitudes in the visual can be seen if the dust opacity is accounted for in the a posteriori radiative transfer (consistent with the dynamical modelling). Increasing with decreasing wavlengths this dimming reflects the characteristic absorption of amC grains shown in Fig.\,\ref{f:specdusteffects} (inset of the bottom panel). This fundamental behaviour is valid for every phase as can be concluded from the bars in Fig.\,\ref{f:photdusteffects} depicting the range of variation of the synthetic photometry during the light cycle. While the amplitude in the V-filter only amounts to $\Delta V$\,$\approx$\,1$^{\rm mag}$ if the dust is not accounted for, we find $\approx$\,4$^{\rm mag}$ for the full calculation including dusty effects. Note, that the amplitudes in Fig.\,\ref{f:photdusteffects} are slightly smaller than the ones plotted in Fig.\,\ref{f:amplitudesDMA}, as the former only cover one pulsation cycle while for the latter all three cycles of Fig.\,\ref{f:lightcurvesDMA} were accounted for (cycle-to-cycle variations).

The fact that the dust absorption in the optical is more relevant than the effect of the modified atmospheric structure (more extended, changing with time; cf. Fig.\,\ref{f:structure}) on the resulting molecular features\footnote{It may be the other way round for M-type LPVs where the circumstellar dust is probably more transparent in the visual (e.g., Woitke \cite{Woitk06b}+\cite{Woitk07}, H\"ofner \cite{Hoefn08}, Bladh et al. \cite{BlaHA11}) while the even larger photometric variations over the light cycle in this wavelength region are caused by the pronounced changes in absorption due to the strongly temperature sensitive TiO features (e.g., Lattanzio \& Wood \cite{LattW04}, Reid \& Goldston \cite{ReidG02}, Hron \cite{Hron06}).} -- which could presumably play an important role as well -- is illustrated in a different way in Fig.\,\ref{f:photdusteffects1}. If only the atomic and molecular lines are taken into account the dynamic model varies in this diagram around the hydrostatic model (not being equal at any phase, though) with deviations of approximately $\pm$\,0$\fm$2 in the colour indices. Switching on the dust absorption leads to a quite elongated loop which is offset by more than 1$^{\rm mag}$ in both colours. Although the influence of molecular features are apparent, the effects due to dust are even more prominent.

%********************************************************************
\subsection{Temporal variations}
\label{s:tempvar}

For all the modelling results presented below all opacity sources described -- i.e., atomic and molecular lines as well as dust -- are accounted for.

\begin{figure}
\resizebox{\hsize}{!}{\includegraphics[clip]{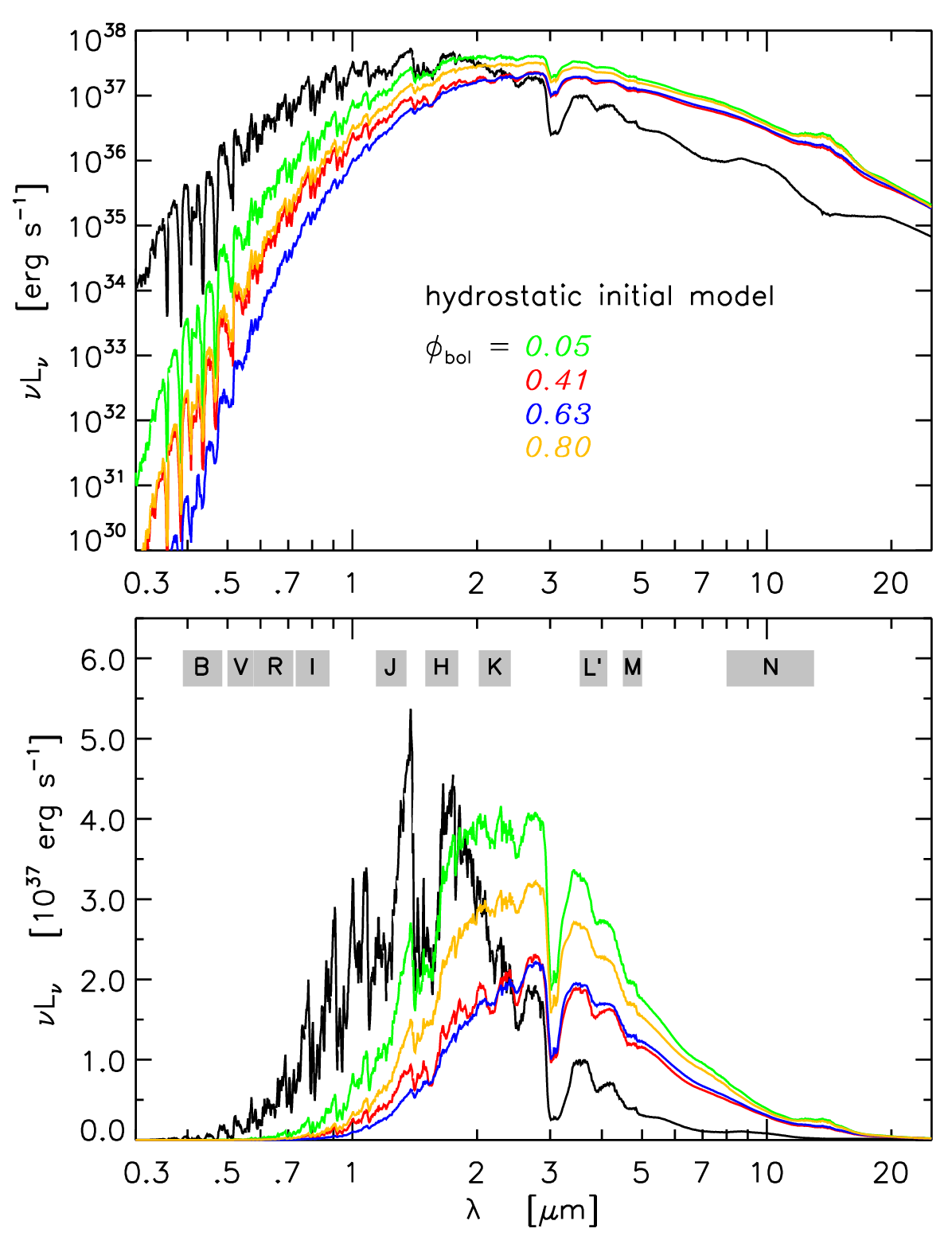}}
\caption{Synthetic low-resolution spectra (plotted on logarithmic and linear scale, respectively) for the whole wavelength range considered based on the initial model as well as different phases of the dynamic calculation of model~S.}
\label{f:specvariationDMA}
\end{figure}

Low-resolution spectra for model~S are shown in Fig.\,\ref{f:specvariationDMA}, illustrating the temporal variations of the spectral appearance of a pulsating C-type Mira with dusty outflow (see Fig.\,A.1 in NHA10 for the corresponding plot of a pulsating model atmosphere without wind). Coupled to the sinusoidally varying luminosity input at the inner boundary of the model (cf. Fig.\,\ref{f:structure}a), we find significant spectral variations during the light cycle. For the radial structures we find in Sect.\,\ref{s:DMAdetails} that the dynamic phases never resemble the hydrostatic atmosphere, a fact that also shows up here for the resulting spectra. The spectrum based on the hydrostatic initial model is different from the corresponding spectra of the dynamic model with the fully developed wind at any phase $\phi_{\rm bol}$. The spectra show a change in the balance of the whole SED ($\phi_{\rm bol}$\,=\,{\it 0.41} vs. {\it 0.63}) and a significantly modified appearance of the molecular features with a pronounced attenuation below 2\,$\mu$m.

\begin{figure}
\resizebox{\hsize}{!}{\includegraphics[clip]{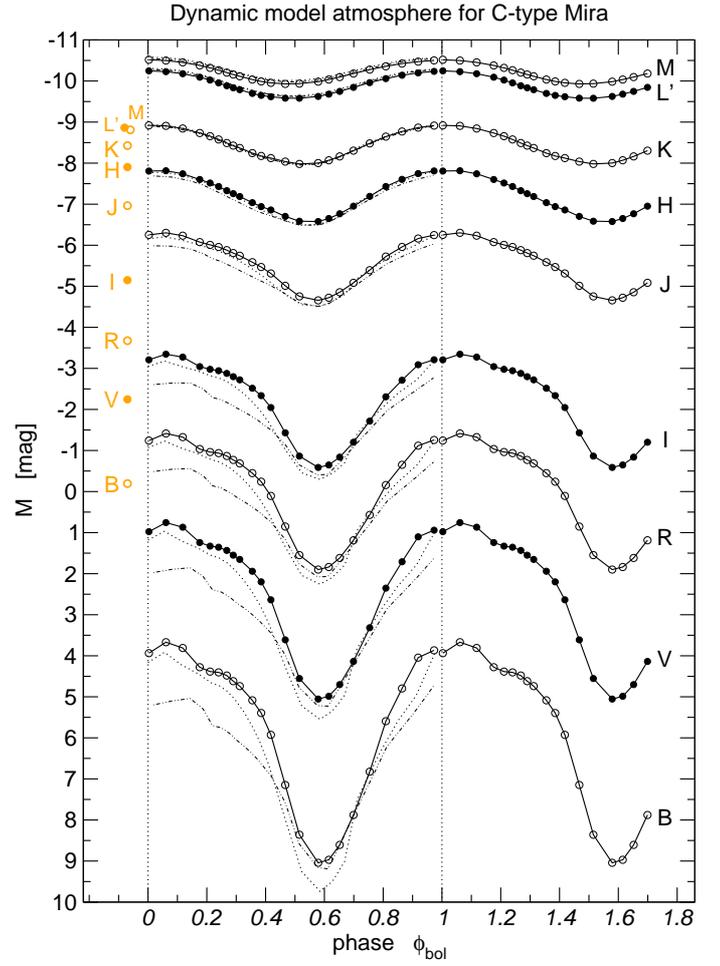}}
\caption{Synthetic lightcurves (absolute magnitudes; $\phi_{\rm bol}$\,=\,[{\it 0,1}]) of model~S computed for various filters as labelled on the right hand side. The data points are repeated to emphasise the periodicity. The corresponding data from two other pulsation cycles ($\phi_{\rm bol}$\,=\,[{\it 7,8}] and [{\it 14,15}]) are overplotted with dotted and dash-dotted lines. On the left hand side, the photometric magnitudes based on the hydrostatic initial model are drawn in for comparison with the same colour code as in Fig.\,\ref{f:specdusteffects}.}
\label{f:lightcurvesDMA}
\end{figure}

\begin{figure}
\resizebox{\hsize}{!}{\includegraphics[clip]{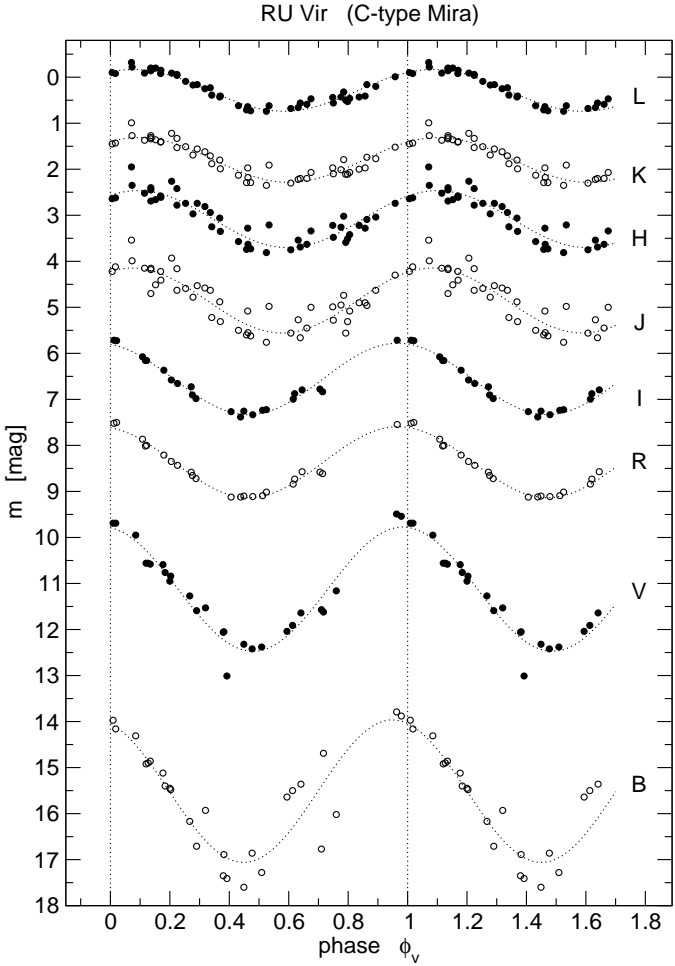}}
\caption{Observed light variations of RU\,Vir (cf. Sect.\,\ref{s:obsresults}) in different filters as labelled on the right hand side. The data were adopted from Eggen (\cite{Eggen75a}; BVRI) and Whitelock et al. (\cite{WhFMG06}; JHKL), respectively. Measurements from different periods are merged into a combined light cycle, each data point is then plotted twice to highlight the periodic variations. Sinusoidal fits (one Fourier component, see text for details) are overplotted with dotted lines to guide the eye. An analogous plot for an \mbox{M-type} Mira, namely RR\,Sco, can be found in Lattanzio \& Wood (\cite{LattW04}; their Fig.\,2.44).}
\label{f:lightcurvesRUVir}
\end{figure}

In Fig.\,\ref{f:lightcurvesDMA} the photometric variations in the chosen standard broad-band filters for the three pulsation cycles of model~S we analysed are shown. As can be seen, the dynamic model becomes fainter in comparison with the initial model for the filters BVRIJH, while it appears brighter for the filters L$^\prime$M. In the \mbox{K-band,} the result of the dynamic calculation oscillates around the hydrostatic magnitude. The corresponding photometric data for RU\,Vir are shown in Fig.\,\ref{f:lightcurvesRUVir}, where the observational data are plotted co-phased (cf. description in Sect.\,\ref{s:obsresults}) and repeated for better illustration. A comparison of the synthetic lightcurves with the observed time-series photometry reveals that our model describes the basic behaviour of C-type Miras\footnote{assuming RU\,Vir to be typical} reasonably well. The continuous increase in brightness with longer wavelengths (from B to L) is reproduced; note the quite similar range in magnitudes on the ordinates. We also find similar absolute magnitudes compared to the model if the apparent magnitudes of RU\,Vir are shifted by the distance modulus listed in Table\,\ref{t:RUVirparameters}. 

Since the observed photometry is adopted from different light cycles and then combined, the scatter around the sinusoidal fits in Fig.\,\ref{f:lightcurvesRUVir} reflects cycle-to-cycle variations. For the model one can recognise such differences between different cycles, especially in the visual and red, directly from the overplotted individual lightcurves in Fig.\,\ref{f:lightcurvesDMA}. The cycle-to-cycle variations are more distinct towards shorter wavelengths and may be attributed to dust formation of varying intensity during the different cycles of the model. 

The observations of RU\,Vir show the known trend of IR lightcurves lagging behind the visual lightcurves by $\approx$\,0.1\,--\,0.2 in phase (see also the extensive discussion concerning this topic in NHA10, App.\,A). For the modelling results we find a phase shift the other way round, with light minima in the visual detectable later in phase by $\Delta \phi_{\rm bol}$\,$\approx$\,{\it 0.1}. In NHA10 we found the same effect based on a different dynamic model atmosphere (model~M), and we showed that both, the amplitude and the shape of the visual light curve are dominated by dust opacities. The distinctive attenuation of molecular bands in the visual and red wavelength range of our current synthetic spectra for phases near to or shortly after the light minimum (Fig.\,\ref{f:specvariationDMA}) fits into this scenario. For model~M in NHA10, the phase of the visual minimum is closely linked to the emergence of a new dust layer, whereas the NIR light curves are close to the (sinusoidal) bolometric light curve (determined by the variable luminosity at the inner boundary), both in shape and phase. Therefore, the discrepancies in phase lags between the visual and NIR compared to observations most likely indicate problems with the dust description in the models, which has two basic aspects: (i) the time-dependent formation and growth of grains, described by the so-called moment method, and (ii) the grain opacities, described by the small particle limit of Mie theory. Dynamic models indicate that typical grain sizes in C-rich AGB stars may exceed the range where the widely-used small particle limit is a good approximation (e.g. Winters et al. \cite{WiFLS97}, Mattsson et al. \cite{MatWH10}, Mattsson \& H{\"o}fner \cite{MattH11}) which may have consequences for the synthetic spectral energy distributions and light curves. A detailed discussion of this point will be presented in a future paper.

\begin{figure}
\resizebox{\hsize}{!}{\includegraphics[clip]{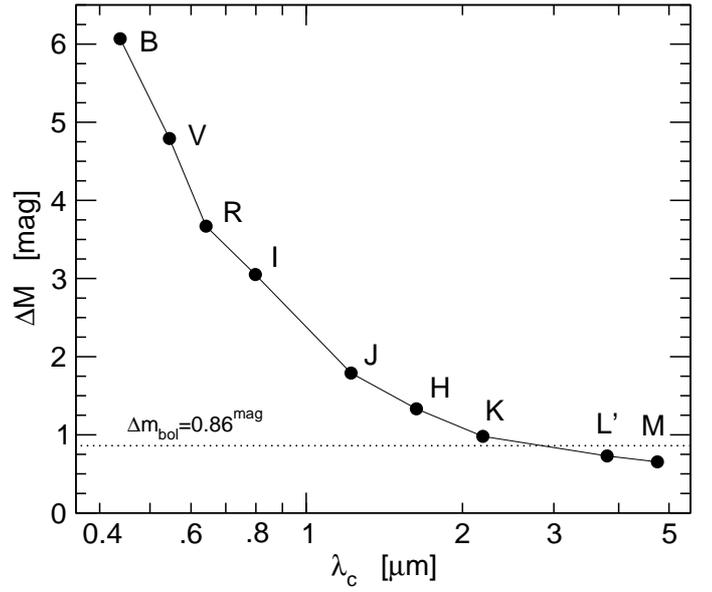}}
\caption{Amplitudes of the synthetic lightcurves for model~S as shown in Fig.\,\ref{f:lightcurvesDMA} (maximum value from the three different cycles minus the corresponding minimum value) plotted against the central wavelengths of the different filters. The amplitude of the variation in $m_{\rm bol}$ of model~S (cf. Fig.\,\ref{f:structure}) is marked, too.}
\label{f:amplitudesDMA}
\end{figure}

\begin{figure}
\resizebox{\hsize}{!}{\includegraphics[clip]{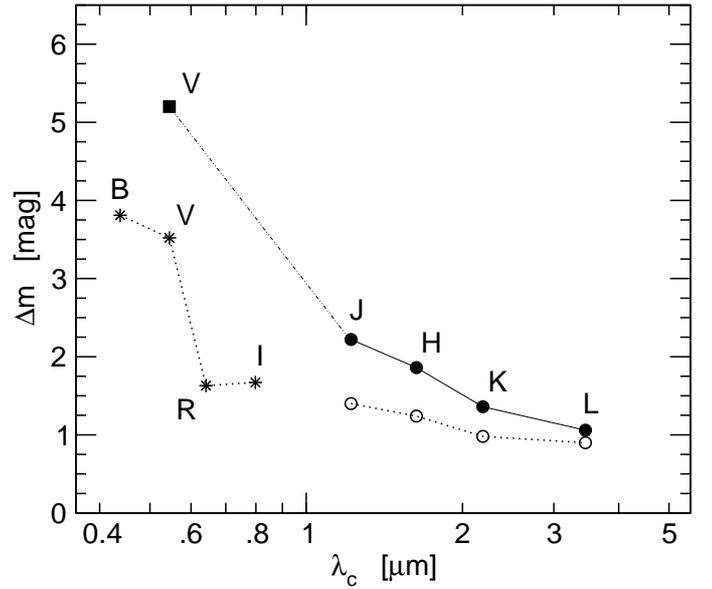}}
\caption{Amplitudes of the observed light variations of RU\,Vir as shown in Fig.\,\ref{f:lightcurvesRUVir}. For the JHKL filters the total amplitudes ($m_{\rm max}$\,--\,$m_{\rm min}$) of the data sample from Whitelock et al. (\cite{WhFMG06}) are plotted with filled circles, together with the amplitude of the Fourier fit given by Whitelock et al. in their Table\,3 plotted with open circles. For the BVRI filters only the first kind of amplitude is available and plotted with asterisks. In addition, the GCVS value for $\Delta V$ is shown with a filled square. The discrepancy in $V$ is due to the limited sampling of the data obtained by Eggen (\cite{Eggen75a}).}
\label{f:amplitudesRUVir}
\end{figure}

From observational studies of evolved LPVs it is known that the variability is, in general, strongest in the blue/visual (e.g., Eggen \cite{Eggen75b}, Mattei \cite{Matte97}, Mennessier et al. \cite{MenBM97}) and slightly less  pronounced in the red (e.g., Fig.\,5 in Hughes \& Wood \cite{HughW90}, Lockwood \& Wing \cite{LockW71}),  while the photometric amplitudes decrease towards the NIR (e.g., Fig.\,7 in Le\,Bertre \cite{LeBer92} and Le\,Bertre \cite{LeBer93}, Percy et al. \cite{PMNPS08}). This fact is also apparent for RU\,Vir in Fig.\,\ref{f:lightcurvesRUVir}. The dynamic model atmosphere used for our modelling reproduces the characteristic decrease of amplitude with wavelength\footnote{This general qualitative trend (mainly determined by the dust effects; cf. Fig.\,\ref{f:specvariationDMA} or Fig.\,A.5 in NHA10) was already reproduced by the models of Winters et al. (\cite{WiFGS94}, \cite{WiFLS97}). Note, however, that their grey models for LPVs with heavy mass loss have optically thick dust shells and Winters et al. (\cite{WiFLS97}) could only compare synthetic lightcurves with IR data of dust-enshrouded stars which are very faint in the optical. The dynamic model atmospheres applied here are also suitable to describe Miras of intermediate mass loss where the inner dust-free photosphere can be observed spectroscopically (molecular features). This allows a comparison of the lightcurves from the visual to the IR (Fig.\,\ref{f:lightcurvesDMA}) with the corresponding data of objects for which the light variations in the optical can be observed (Fig.\,\ref{f:lightcurvesRUVir}). \rm} quite well as can be seen in Fig.\,\ref{f:lightcurvesDMA}. A direct  comparison of the photometric amplitudes in a quantitative way can be drawn from Figs.\ref{f:amplitudesDMA} and \ref{f:amplitudesRUVir}. In the observational plot we also included the value for $\Delta V$ from the GCVS catalogue (Table\,\ref{t:RUVirparameters}), as the rather limited time-series of photometric data in the visual obtained by Eggen (\cite{Eggen75a}) may not cover the full amplitude as it can be found for studies over a large number of light cycles (cf. Fig.\,2b in Mattei \cite{Matte97}). The reason for this are the known cycle-to-cycle variations which are also responsible for the Fourier amplitudes always being smaller than the total amplitudes of individual measurement ($m_{\rm max}$\,--\,$m_{\rm min}$). The fact that the Fourier fit underestimates the amplitude can be recognised in Fig.\,\ref{f:amplitudesRUVir} for the NIR filters. Our model also shows the correlation $\Delta K$\,$\approx$\,$\Delta m_{\rm bol}$ (Fig.\,\ref{f:amplitudesDMA}) which was found observationally for Miras by Whitelock et al. (\cite{WhiMF00}; their Fig.\,5), although this was investigated only for \mbox{O-rich} targets with moderate MLRs.

\begin{figure}
\resizebox{\hsize}{!}{\includegraphics[clip]{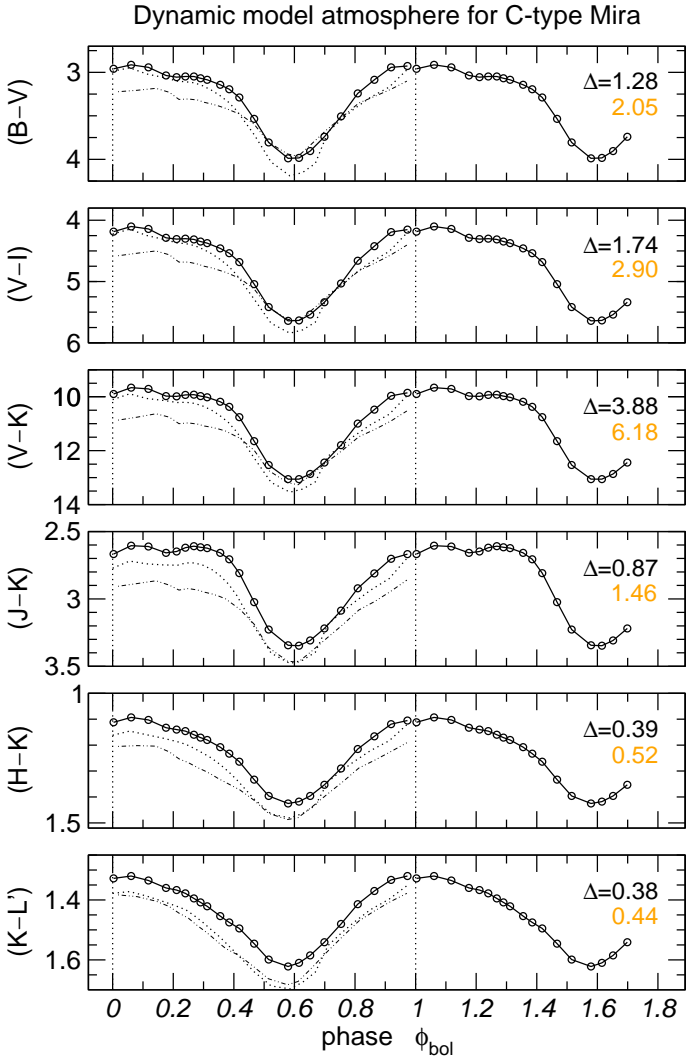}}
\caption{Variations of different colour indices for model~S, overplotted with the same linestyles as in Fig.\,\ref{f:lightcurvesDMA} are the values for different pulsation cycles. On the right hand side the amplitudes in the colour indices are listed as well as the corresponding colours of the hydrostatic initial model for comparison. Same colour code as in Fig.\,\ref{f:specdusteffects}.}
\label{f:colorcurvesDMA}
\end{figure}

\begin{figure}
\resizebox{\hsize}{!}{\includegraphics[clip]{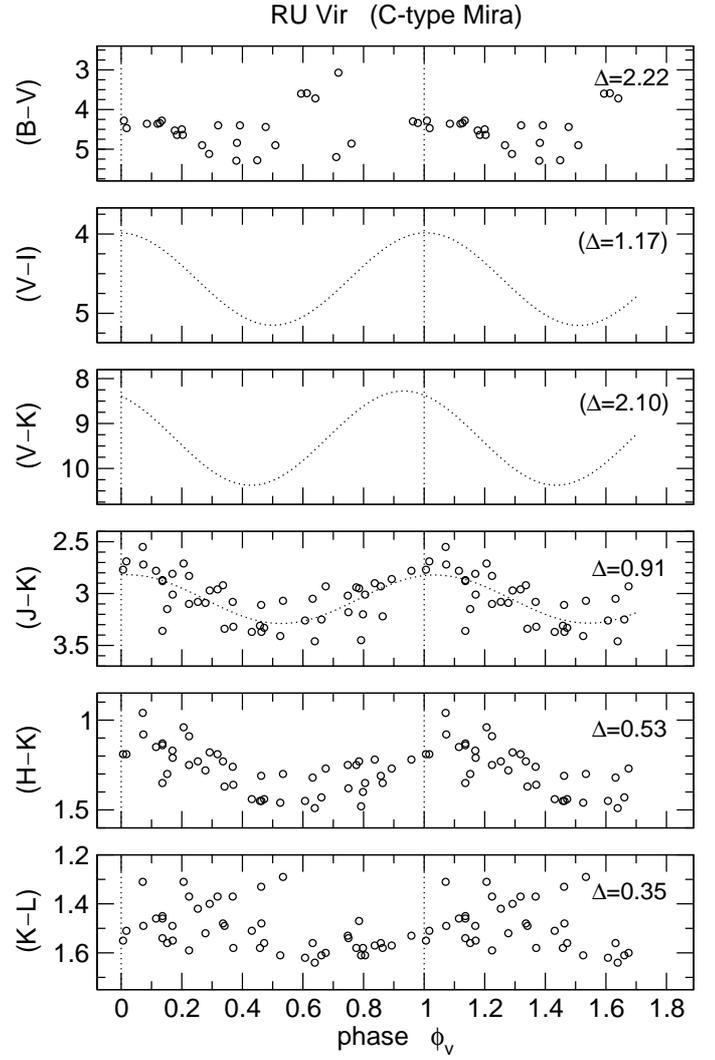}}
\caption{Variations of different colour indices of RU\,Vir. The dotted curves were derived from the sinusoidal fits in Fig.\,\ref{f:lightcurvesRUVir} as no simultaneous measurements of the individual magnitudes are available for ($V$\,--\,$I$) and ($V$\,--\,$K$). To get an idea of how representative these estimates are the same simulated colour curve was derived for ($J$\,--\,$K$), where photometry at the same epochs is available as well. On the right hand side the amplitudes of the respective colour indices are listed (in brackets if only derived from the simulated variations).}
\label{f:colorcurvesRUVir}
\end{figure}

Figure\,\ref{f:colorcurvesRUVir} shows the variations of RU\,Vir in various colours based on the photometry shown in Fig.\,\ref{f:lightcurvesRUVir}. The known trend (e.g., Payne-Gaposchkin \& Whitney \cite{PaynW76}, Lebzelter \& Posch \cite{LebzP01}) of bluer colours around light maximum and redder colours around minimum can be recognised, at least for selected colour indices, such as ($J$\,--\,$K$). Still, we are facing a severe scatter, resulting from the merging of measurements from very different periods into one combined light cycle. The variation can be seen much clearer for ($V$\,--\,$I$) and ($V$\,--\,$K$), where we derived \textit{simulated} colour curves (dotted lines) from the sinusoidal lightcurve fits of Fig.\,\ref{f:lightcurvesRUVir} as no simultaneous measurements of the individual magnitudes are available for these colour indices. However, the amplitude in these colour indices has to be regarded as a lower limit (see discussion above). This is illustrated in the panel for \mbox{($J$\,--\,$K$)} of Fig.\,\ref{f:colorcurvesRUVir}, where we overplotted the original data with the same kind of simulated variation. While the amplitude \mbox{$\Delta$($J$\,--\,$K$)} of the simultaneously obtained data amounts to 0$\fm$91, the simulated colour curve shows an amplitude of only 0$\fm$47. The corresponding temporal variations in colour indices for the dynamic model are shown in Fig.\,\ref{f:colorcurvesDMA}. We find a good agreement with the observations of RU\,Vir in the principal behaviour, i.e., the absolute values and the amplitudes. Comparing for example the ($J$\,--\,$K$) colours in Figs.\,\ref{f:colorcurvesDMA} and \ref{f:colorcurvesRUVir} we find the model resembling RU\,Vir very well. Despite some cycle-to-cycle variations, the model appears clearly bluer around maximum phases $\phi_{\rm bol}$\,$\approx$\,{\it 0.0} and redder at minimum phases $\phi_{\rm bol}$\,$\approx$\,{\it 0.5}.

Noteworthy in Fig.\,\ref{f:colorcurvesDMA} is the difference between the colour indices of the hydrostatic initial model also given there [e.g., \mbox{($J$\,--\,$K$)}\,=\,1.46] and the colour range covered by the dynamical model during a pulsation cycle [e.g., ($J$\,--\,$K$)\,$\approx$\,2.5--3.5]. For the latter we find significantly larger values due to the circumstellar reddening caused by the grains of amC dust in the stellar wind (Sect.\,\ref{s:opacsources}). This fact provides the explanation for the discrepancies identified in Paper\,I. There we found distinct deviations in various colour indices between the hydrostatic COMARCS models and observed carbon stars (cf. Figs.\,14, 15, 16, and 18 in Paper\,I). The very evolved and cool C-type giants appear much redder than predicted by the COMARCS models, which are found to deliver realistic results only for objects above $T_{\rm eff}$\,$\approx$\,2800\,K. The samples of observed stars below this approximate border have, for example, ($J$\,--\,$K$) colours larger than 2$^{\rm mag}$ as can be seen in Figs.\,14 and 18 of Paper\,I. While the hydrostatic models there show the discussed colour reversion effect with decreasing effective temperatures which leads to ($J$\,--\,$K$) colours of $<$\,1.6$^{\rm mag}$ (similar to the hydrostatic initial model analysed here), the dynamical model exhibits ($J$\,--\,$K$) colours redder by at least 1$^{\rm mag}$. Thus, the circumstellar reddening results in colour indices much closer to observed values of stars where the assumptions for the COMARCS models (hydrostatic configuration, no circumstellar dust) are not adequate anymore. In other words, the dynamic model atmospheres with dusty outflows -- as model~S, studied in this work -- represent the only self-consistent method to describe (colours of) mass-losing LPVs. This is even more true if we proceed from the sample of quite moderate carbon stars of Bergeat (\cite{BerKR01}) used in Paper\,I to the sample of more evolved (redder, higher MLRs) \mbox{C-type} Miras of Whitelock et al. (\cite{WhFMG06}) used for comparison with our modelling results below (e.g., Fig.\,\ref{f:JHKObsMAERCSE}).

\begin{figure}
\resizebox{\hsize}{!}{\includegraphics[clip]{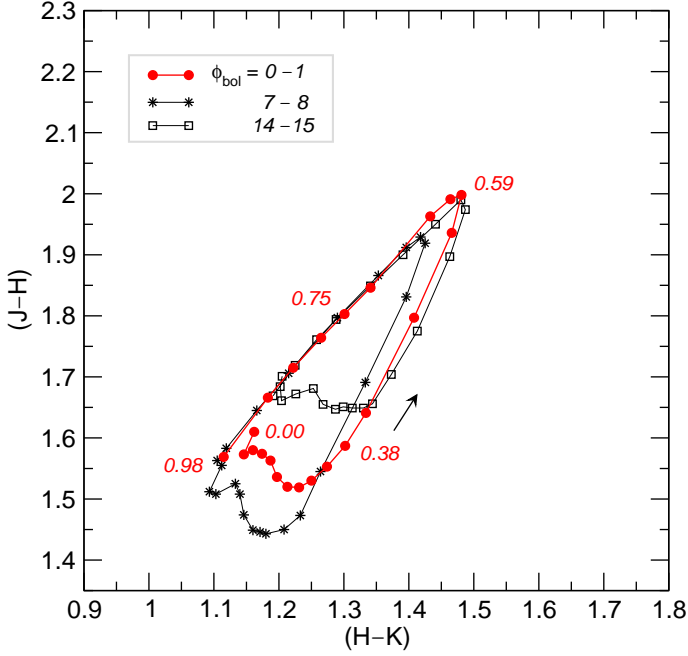}}
\caption{Colour-colour diagram for model~S where the values for the three different pulsation cycles considered are plotted. For orientation purposes the arrow illustrates how the model loops through the diagram during one cycle and selected phases are labelled.}
\label{f:JHKDMA}
\end{figure}

%********************************************************************
\subsection{Colour-colour diagrams}
\label{s:CMDCCD}

In the following we want to compare the results of our synthetic photometry with colour-colour diagrams (CCDs) usually investigated in observational studies. We chose one type of CCD which is commonly used in AGB research, namely ($J$\,--\,$H$) vs. \mbox{($H$\,--\,$K$)}. In a first step, we compare model~S with a selected sample of observed C-Miras (among them RU\,Vir) that show similar MLRs and (variations of) colours to demonstrate that the dynamical model provides realistic NIR colour indices. In the second step, we contrast the synthetic colours with the observational data of the whole sample of C-rich Miras from Whitelock et al. (\cite{WhFMG06}), which covers a wider range in parameters (especially MLRs). 

\begin{figure}
\resizebox{\hsize}{!}{\includegraphics[clip]{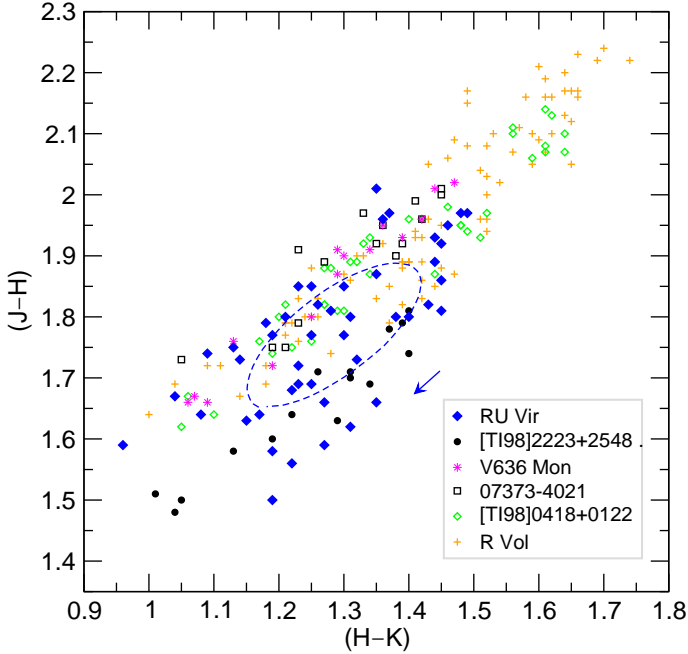}}
\caption{Colour-colour diagram containing observations at several epochs for a sample of C-type Miras adopted from Whitelock et al. (\cite{WhFMG06}) as listed in their Table\,2 (no correction for interstellar reddening). In addition, the dashed line marks the average variation of RU\,Vir during the light cycle as derived from the sinusoidal fits in Fig.\,\ref{f:lightcurvesRUVir}. The arrow marks the direction of how the star passes through the loop.}
\label{f:JHKObs}
\end{figure}

In Fig.\,\ref{f:JHKDMA} we plot ($J$\,--\,$H$) vs. \mbox{($H$\,--\,$K$)} for the three different periods of model~S we made use of. A significant variation of approximately half a magnitude in both colours during the light cycle can be recognised. In Fig.\,\ref{f:JHKObs} we show photometric measurements for selected galactic C-type Miras adopted from Whitelock et~al. (\cite{WhFMG06}) for comparison. Several targets from their sample were found to populate the same region (note the same scaling of Figs.\,\ref{f:JHKDMA} and \ref{f:JHKObs}) in this CCD, also the variations over time in the respective colour indices show similar ranges. This is especially true for the reference object RU\,Vir. All of the targets shown in Fig.\,\ref{f:JHKObs} have MLRs of $\approx$\,1--3\,$\times$\,10$^{-6}$\,$M_{\odot}\,$yr$^{-1}$ (Whitelock et~al. \cite{WhFMG06}), which is quite similar to the corresponding value of the model (Table\,\ref{t:dmaparameters}). Eye-catching is the fact that the model loops through the ($J$\,--\,$H$)-($H$\,--\,$K$)-plane once every pulsation cycle, due to the phase-dependent dust absorption and the different behaviours of molecular features contributing to the emerging spectra at the respective wavelengths within different filters. In the model (see the loops with different plot symbols and colours), we also found some cycle-to-cycle variations, especially for phases around light maximum. This reflects the fact, that atmospheric structure and dust formation can differ severly for the same phases $\phi_{\rm bol}$ of different cycles of the model.

It is not possible to investigate the corresponding variations as a function of phase for RU\,Vir directly as the individual data points in Fig.\,\ref{f:JHKObs} were obtained over several periods of the star with an insufficient sampling rate (Sect.\,\ref {s:obsresults}, Fig.\,\ref{f:lightcurves}). However, we derived simulated variations based on the Fourier-fits of the lightcurves (Fig.\,\ref{f:lightcurvesRUVir}), just as it was done for Fig.\,\ref{f:colorcurvesRUVir}. This results in a loop -- also shown in Fig.\,\ref{f:JHKObs} (dashed line) -- similar to the model, with $\phi_{\rm bol}$\,$\approx$\,{\it 0.0} at the lower left of the ellipse and $\phi_{\rm bol}$\,$\approx$\,{\it 0.5} at the upper right end. The range covered by the loop is again smaller than the corresponding range of the individual measurements as the fits tend to underestimate the variations (cf. Sect.\,\ref{s:tempvar}). We noticed that while the model loops counter-clockwise, the average variation of RU\,Vir follows a clockwise movement.\footnote{This is related to the fact that the phase shifts between lightcurves at different wavelengths show opposite trends in the model compared to observations (Sect.\,\ref{s:tempvar}). See also App.\,\ref{s:loopsense}.} Hints for the latter were also found by other observational studies in the past (Fig.\,13 in Smak \cite{Smak64}, Payne-Gaposchkin \& Whitney \cite{PaynW76}, Fig.\,2 in Alvarez \& Plez \cite{AlvaP98}). A similar disagreement between dynamic models of M-type objects (with a parameterised description of dust formation) and real stars was also noticed by Lebzelter et al. (\cite{LNHLH10}; their Fig.\,20) in the context of equivalent width variations of spectral lines over the light cycle. A more detailed investigation of this effect in the future may reveal interesting insights into the adequateness of certain modelling assumptions (e.g. the inner boundary condition simulating the stellar pulsation).

\begin{figure}
\resizebox{\hsize}{!}{\includegraphics[clip]{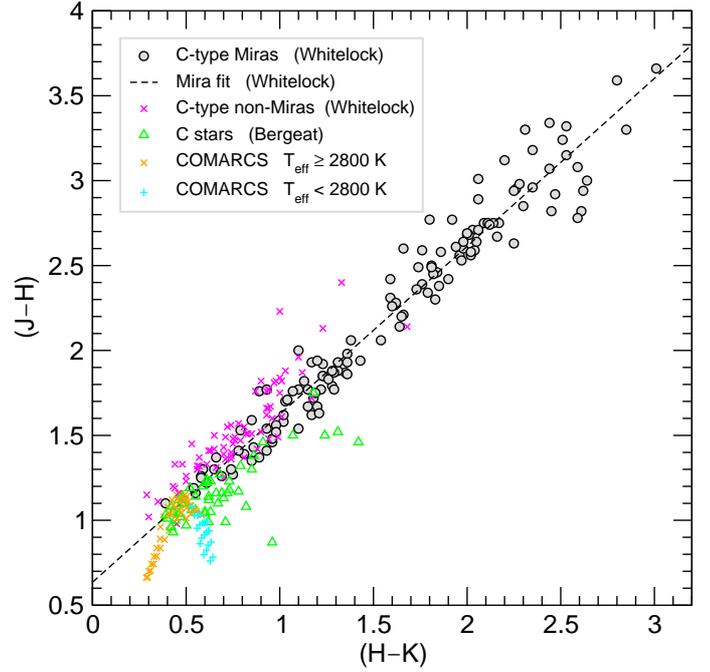}}
\caption{Mean colours (no correction for interstellar reddening) for C~stars derived from the Fourier-mean JHK magnitudes listed in Table\,3 of Whitelock et al. (\cite{WhFMG06}) for all Miras (i.e., their variability type 1n) and all non-Miras (i.e., their class 2n). Each point represents the time average of temporal variations (as shown in Fig.\,\ref{f:JHKObs} for a subsample of the targets included here) for a specific object. The dashed line represents the fit given in Eq.\,7 of Whitelock et al. (\cite{WhFMG06}) for the Mira sample prior to de-reddening. In addition, the photometric data collected by Bergeat et al. (\cite{BerKR01}; Table\,4) is overplotted as well as synthetic colours from a sub-grid of the hydrostatic COMARCS models presented in Paper\,I for comparison.}
\label{f:JHKObsMAERCSE}
\end{figure}

To put the selected dynamic model into a broader context, we compiled further observational data for the same kind of CCD. This allowed to investigate the position of model~S (as well as RU\,Vir) on an empirical MLR-sequence. Figure\,\ref{f:JHKObsMAERCSE} shows a compilation of literature data for a wider range in both of the colour indices compared to Fig.\,\ref{f:JHKObs}. Note that while in the latter individual measurements at different epochs are shown, the colours plotted now in Fig.\,\ref{f:JHKObsMAERCSE} represent mean values and every point stands for one object. For the targets from Whitelock et al. (\cite{WhFMG06}) we used the data prior to dereddening, enabling us to include also stars with small photometric variations (non-Miras). For illustration purposes there are also measurements for carbon stars from Bergeat et al. (\cite{BerKR01}) as well as synthetic photometry based on a number of hydrostatic model atmospheres from Paper\,I shown. Despite the minor inconsistency that the latter colours are not subject to interstellar reddening, the comparison in Fig.\,\ref{f:JHKObsMAERCSE} is useful to demonstrate the influence of pulsation and dust effects on colour indices in general. Late-type giants with C-rich atmospheric chemistry can be found over quite a range in NIR colours. However, the hydrostatic models (at least the relevant sub-sample with temperatures above 2800\,K) are only able to describe the observations on the very blue end of the sequence where we find objects exhibiting rather moderate photometric variations (or none at all) and low mass loss rates (or none at all). Both of these properties become more and more pronounced if one proceeds to the upper right corner of this CCD. There we find distinct Mira variables with much redder colours caused by the significant mass loss of this objects (cf. Fig.\,16 of Whitelock et al. \cite{WhFMG06}). 

\begin{figure}
\resizebox{\hsize}{!}{\includegraphics[clip]{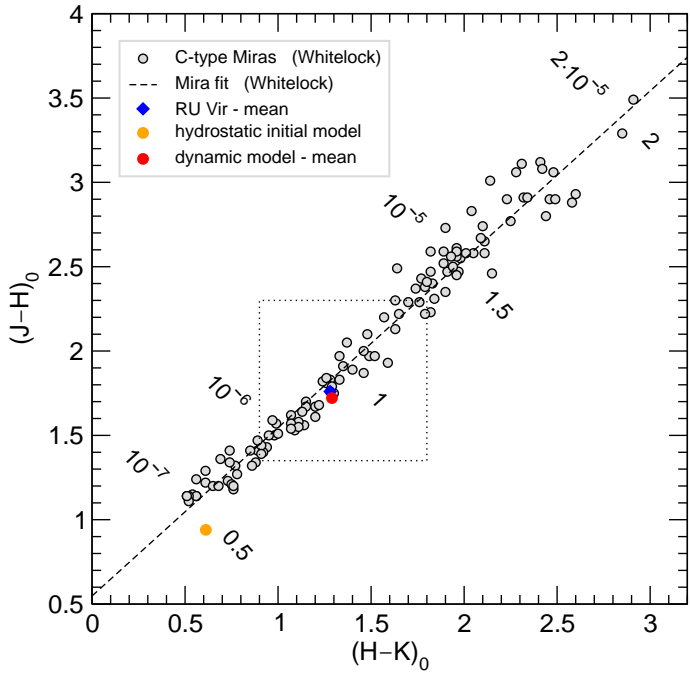}}
\caption{Mean colours corrected for interstellar reddening as listed in Table\,6 of Whitelock et al. (\cite{WhFMG06}) for all C-type Miras (i.e., their class 1n). The dashed line represents the fit given in Eq.\,2 of Whitelock et al. (\cite{WhFMG06}) for the de-reddened Mira sample. The labels along the sequence mark approximate mass loss rates $\dot M$~[$M_{\odot}\,$yr$^{-1}$] (above the fit line) as well as photometric amplitudes $\Delta K$~[mag] (below) corresponding to given colours as derived by Whitelock et al. Overplotted are the synthetic colours of model~S (initial model as well as the average of several phases of the various cycles in Fig.\,\ref{f:JHKDMA}) for comparison. The box drawn with dotted lines marks the range covered in Figs.\,\ref{f:JHKDMA}+\ref{f:JHKObs}.}
\label{f:JHKObsDMA}
\end{figure}

This fact is demonstrated in Fig.\,\ref{f:JHKObsDMA} in a more quantitative way by using only the Mira subsample of Whitelock et al. (\cite{WhFMG06}), for which these authors derived and listed useful quantities (esp., dereddened colours, amplitudes of photometric variations, MLRs). With its intermediate mass-loss rate and an photometric amplitude of $\Delta K$\,=\,0$\fm$98 (Table\,\ref{t:RUVirparameters}), RU\,Vir is located somewhat in the middle of the sequence of increasing MLRs and $\Delta K$ which can be recognised in Fig.\,\ref{f:JHKObsDMA}. The same applies also for model~S (with an $\langle\dot M\rangle$ of $\approx$\,4.3\,$\times$\,10$^{-6}$$M_{\odot}\,$yr$^{-1}$), the average synthetic colour of which falls nicely onto the observed sequence for C-type Miras. Also the amplitude of model~S ($\Delta K$\,=\,1$^{\rm mag}$) fits quite well to what is observed for targets populating the same region in the CCD of Fig.\,\ref{f:JHKObsDMA}. This basic finding is independent of whether the interstellar dereddening was accounted for (Fig.\,\ref{f:JHKObsDMA}) or not (Fig\,\ref{f:JHKObsMAERCSE}). The difference between the mean colour of the dynamical model and the corresponding value based on the initial model  illustrates again the effect of circumstellar reddening. For an easier comparison of the full sequence of observed Miras with the temporal variations of model~S in Fig.\,\ref{f:JHKDMA}, we also marked the smaller range of this plot in Fig.\,\ref{f:JHKObsDMA}.

%####################################################################
\section{Summary}

The hydrostatic and dust-free COMARCS model atmospheres presented in Paper\,I are successful in reproducing the photometric properties of carbon-rich giants with effective temperatures down to approximately 2800\,K. However, this type of models is only of limited use for the interpretation of observations of more evolved C-rich AGB stars. For such objects dynamic effects -- namely the pulsating stellar interior as well as the development of a stellar wind leading to a circumstellar dusty envelope -- become significant. To cover these we need to apply more sophisticated atmospheric models, including atmospheric dynamics.

In this work we studied the influence of pulsation and circumstellar dust on the photometric appearance of late-type giants with the help of one selected dynamic model atmosphere (model~S) representing a typical C-type Mira losing mass at an intermediate rate. For such models the wind is driven by radiation pressure on amorphous carbon dust particles forming in the cool outer atmospheric layers. For various points in time we carried out detailed RT calculations based on the corresponding atmospheric structures given by the dynamic model. The resulting synthetic spectra were convolved with filter transmission curves of standard broad-band filters in the visual and red (Johnson-Cousins UBVRI) as well as in the NIR (Johnson-Glass JHKL$^\prime$M). By applying adequate zeropoints we obtained synthetic photometry for the model as function of phase during several pulsation cycles.

The modelling of photometric properties can be used to study effects which are not accessible to observations. We investigated, for example, the influence of different opacity sources on the SED. Thereby, we found that the effect of the much more extended atmosphere of the fully developed model (wind) on the prominent molecular features is comparatively low (still apparent, though) when we look at SEDs based on the photometry in the complete set of chosen filters. On the other hand, the characteristic absorption of the amC grains gives rise to pronounced circumstellar reddening and a significant change of the SED. The outer layers of mass-losing LPVs are qualitatively different from hydrostatic and dust-free stars because of the dynamic processes and the dust formation taking place. This fact is reflected clearly in the radial structures, the resulting spectra, and the global SEDs as described by the photometry. Once the models are successfully tested against observations, they are also useful for predicting light variations in filters where no observed data are available or for contributing to population synthesis studies.

Contrasting the synthetic photometry with observational results from the literature provided important and crucial information. Especially, the C-type Mira variable RU\,Vir turned out to be a useful reference object for several reasons (photometric data in several filters available, similar to the dynamical model, etc.). Comparing the observed BVRIJHKL lightcurves of this target (Fig.\,\ref{f:lightcurvesRUVir}) with the corresponding photometric variations of the model (Fig.\,\ref{f:lightcurvesDMA}) revealed good agreement concerning several aspects: 
\begin{itemize}
\item [-]
the objects become, in general, brighter in magnitudes when scanning the filters with increasing central wavelengths (i.e., from B to L) 
\item [-]
the range in magnitudes covered by this sequence,
\item [-]
absolute magnitudes when adopting literature values for the distance modulus of RU\,Vir,
\item [-]
occuring cycle-to-cycle variations, and
\item [-]
the trend of photometric amplitudes decreasing from the visual to the IR.
\end{itemize}

The circumstellar reddening caused by the dusty envelope leads to considerably larger values for all kinds of colour indices when compared to the corresponding data based on hydrostatic (COMARCS or initial) models. Therefore, the dynamic model atmospheres are more appropriate to describe evolved AGB stars. The applied model~S is able to reproduce absolute values and variations of diverse colour indices of RU\,Vir quite well. 

We also tested the model against observations by looking at NIR two-colour diagrams. We found several observed Carbon-Miras showing variations in the ($J$\,--\,$H$)-($H$\,--\,$K$)-plane which are very similar to the synthetic photometry in terms of approximate regions and range of variations during the light cycle. Both, the model and RU\,Vir describe loops in such CCDs, although the orientation turned out to be contrary. Average colour indices in the same diagram compiled for various observed C stars show the known trend of redder colours with increasing mass loss rates. Losing mass at an intermediate rate, RU\,Vir can be found roughly in the middle of the linear sequence crossing the ($J$\,--\,$H$) vs. ($H$\,--\,$K$) diagram. Model~S has almost the same mean colours and, thus, fits very well into this global scenario.

Notwithstanding these successes of our modelling, some discrepancies become apparent from a more detailed comparison of the synthetic photometry with observational data (e.g. shapes of lightcurves, phase lags between visual and IR lightcurves, senses of rotation of loops described in CCDs). These are certainly shortcomings of the current models. However, such differences emerging from a strictly consistent modelling (as carried out in this work) can also provide unique insights into the physics taking place in the outer layers of evolved red giants which will help to improve the modelling approach.

The dynamic model atmospheres applied in this paper were already successful in reproducing observed mass loss properties ($\dot M$, $u_\infty$; Mattsson et al. \cite{MatWH10}), observed SEDs and low-resolution spectra (H\"ofner et al. \cite{HoGAJ03}, Gautschy-Loidl et al. \cite{GaHJH04}), line profile variations in observed high-resolution spectra (Nowotny et al. \cite{NAHGW05}, \cite{NoLHH05}, \cite{NowHA10}), variations in line equivalent widths over the light cycle (Lebzelter et al. \cite{LNHLH10}) and interferometric measurements (Sacuto et al. \cite{SAHNP11}). The synthetic photometry presented here provides a further clue that the models resemble the global atmospheric structures (and their temporal variations) of mass-losing C-type LPVs to a rather reasonable degree. 

In the future, we plan to extend our efforts to model photometric properties and variations -- carried out exemplarily and in detail for one selected atmospheric model, here -- to the grid of dynamical models published by Mattsson et al. (\cite{MatWH10}). Covering a wide range of parameters, this will enable us to test the influence of different parameters (e.g., MLR) on the resulting synthetic photometry.

%####################################################################
\begin{acknowledgements}
We are thankful to M.S.~Bessell for supporting us with his sound knowledge concerning the use of different photometric systems in the past decades (O.J.~Eggen) and for providing the photometric transformation given in Sect.\,\ref{s:obsresults}. 
We thank T.~Lebzelter for his assistance in the use of the program {\tt Period04} to prepare the observational data of RU\,Vir (determination of visual phases) and compute fits of the lightcurves. 
Sincere thanks are given to J.~Hron and T.~Lebzelter for careful reading and fruitful discussions.
This work was supported by the \textit{Fonds zur F\"orderung der Wis\-sen\-schaft\-li\-chen For\-schung} (FWF) under project numbers P18939--N16, P19503--N16 and P21988-N16 as well as the Swedish Research Council. 
BA acknowledges funding by the contracts ASI-INAF I/016/07/0 and ASI-INAF I/009/10/0.
We acknowledge with thanks the variable star observations from the AAVSO International Database contributed by observers worldwide and used in this research.
This research has made use of (i) NASA's Astrophysics Data System, and 
(ii) the SIMBAD database, operated at CDS, Strasbourg, France. 
\end{acknowledgements}

%####################################################################

\Online

\begin{appendix}

\section{Details of the atmospheric models}

%********************************************************************
\subsection{General remarks}

The two aspects discussed in this paper -- pulsation of the stellar interior and the formation of a dusty stellar wind -- have a considerable influence on the outer layers of an AGB star and hydrostatic model atmospheres are, therefore, often not an adequate approach to describe the resulting complex stratifications and dynamic effects. As a consequence, dynamic model atmospheres were developed to simulate mass-losing LPVs (Sect.\,\ref{s:intro}).

The majority of the AGB stars found are of spectral type~M (e.g., Nowotny et al. \cite{NoKOS03}, Battinelli et al. \cite{BatDL03}, Rowe et al. \cite{RoRBC05}) and have oxygen-rich\footnote{In fact this does not mean O-\textit{enriched}, but rather the standard value for abundances of elements in the cosmos.} atmospheric chemistries with C/O\,$<$\,1. This is also reflected in the mineralogical species constituting the dust grains that occur in the winds. Observational spectroscopic studies revealed a rich mineralogy in the dusty outflows of M-type giants (e.g., Molster \& Waters \cite{MolsW03} and references therein). However, the dust formation process in the atmospheres of such objects is so far not fully understood (cf. H\"ofner \cite{Hoefn09}) and the open question concerning the driving mechanism for the wind in the O-rich case is not solved, yet (Woitke \cite{Woitk06b}+\cite{Woitk07}, H\"ofner \cite{Hoefn07}). One possibility to drive a wind was sketched by H\"ofner (\cite{Hoefn08}) with the help of big ($\approx$\,$\mu$m) Fe-free silicate grains (i.e., Forsterite Mg$_2$SiO$_4$) and their substantial radiative scattering cross section.

As a star evolves during the AGB phase, a combination of nucleosynthesis and convection processes can lead to a drastic change of the chemical composition of the outer layers (e.g., Busso et al. \cite{BusGW99}, Herwig \cite{Herwi05}). Having been converted to a carbon star (Wallerstein \& Knapp \cite{WallK98}) with an atmospheric chemistry characterised by C/O\,$>$\,1, the spectrum of this star is strongly changed and shows prominent features of \mbox{C-bearing} molecules (Paper\,I, Fig.\,\ref{f:specdusteffects}). This transformation to spectral type~C is, in addition, highly relevant for the circumstellar dust chemistry which is believed to be simpler than the corresponding one of M-type stars. Apart from the few dust species identifiable via their characteristic emission features, as for example SiC or MgS (see Molster \& Waters \cite{MolsW03}), the majority of the formed dust grains are composed of amorphous carbon (amC). Although they are not recognisable in the spectra by a distinctive spectral feature because of the featureless extinction properties (e.g., Fig.\,\ref{f:specdusteffects} or Andersen et al. \cite{AndLH99}), amorphous carbon dust plays a crucial role for driving stellar winds via radiation pressure as outlined for example in NHA10. Moreover, the formation and evolution of amC grains can be treated in a consistent way (Gail \& Sedlmayr \cite{GailS88}, Gauger et al. \cite{GauGS90}) in numerical models. This enables us to calculate detailed and self-consistent models for the atmospheres of pulsating red giants including the developing winds.

\begin{figure}
\resizebox{\hsize}{!}{\includegraphics[clip]{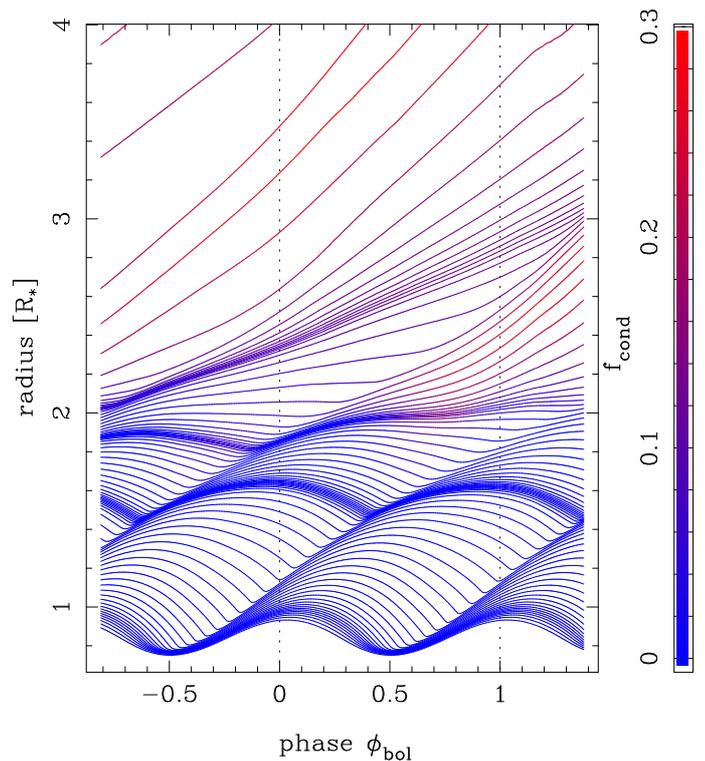}}
\caption{Movement of mass shells with time at different depths of model~S, depicting a pulsation-enhanced dust-driven wind. The shown trajectories follow the evolution with time of certain matter elements starting from the distribution of adaptive grid points (higher density of points at the locations of shocks) at an arbitrarily chosen point in time (in this case at the end of the plotted sequence). Colour-coded is the degree of condensation of the available carbon into amorphous carbon dust grains.}
\label{f:massenschalenS}
\end{figure}

\begin{figure*}
\centering
\includegraphics[width=17cm]{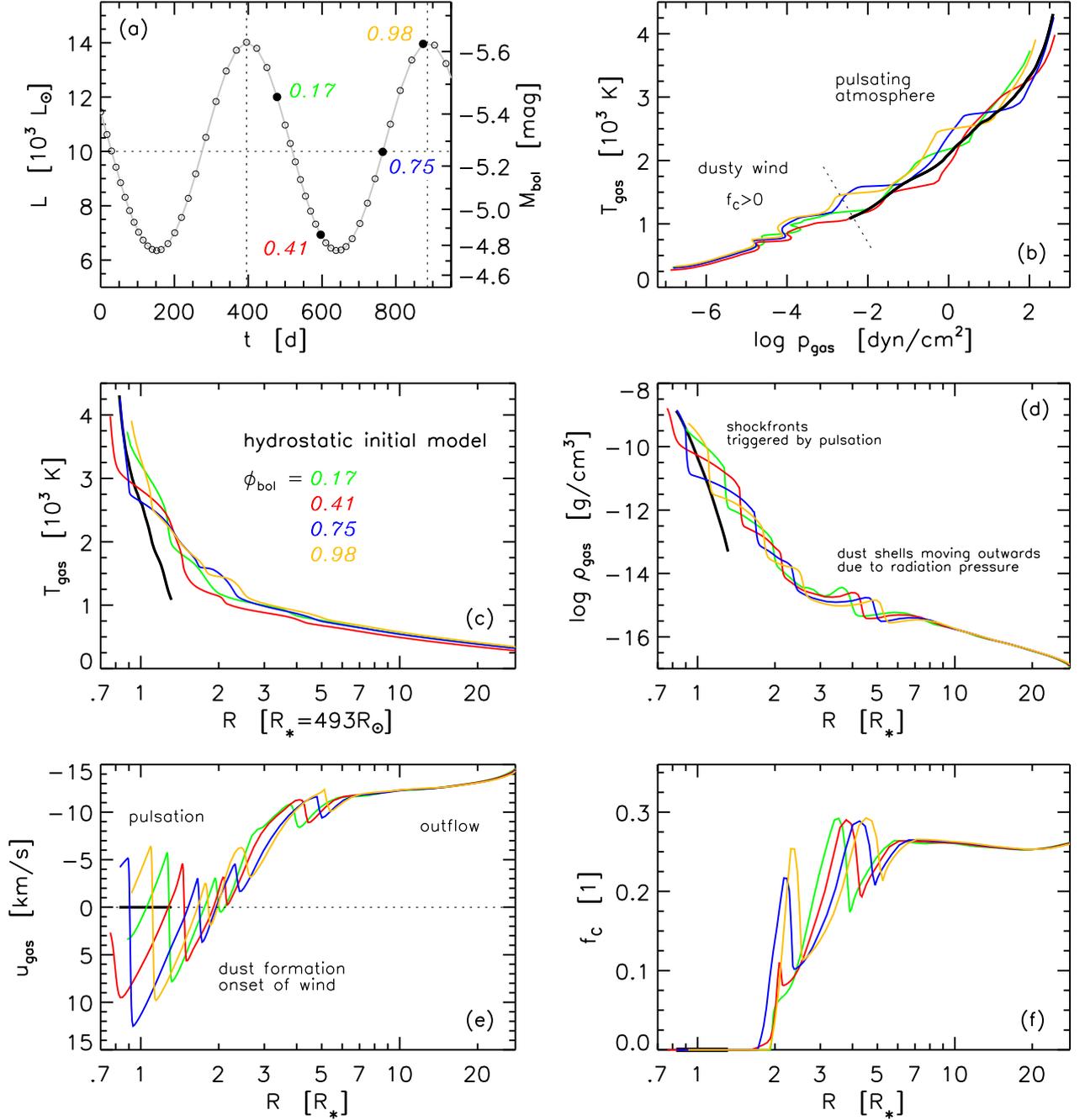}
\caption{
Characteristic properties of model~S. 
Plotted in panel \textbf{(a)} is the bolometric lightcurve resulting from the variable inner boundary (piston), the circles mark instances of time for which snapshots of the atmospheric structure were stored by the radiation-hydrodynamics code, partly labelled with the corresponding values for phase $\phi_{\rm bol}$.
The dotted vertical lines mark the points in time of $\phi_{\rm bol}$\,=\,{\it 0.0} and {\it 1.0}, respectively, while the horizontal line marks $L_\star$ of the initial model which is almost equal to $\langle L\rangle$ of the dynamical calculation (NHA10).
The other panels show the atmospheric structures of the initial hydrostatic model (thick black line) together with selected phases $\phi_{\rm bol}$ of the dynamic calculation during one pulsation cycle (colour-coded in the same way as the phase labels of panel \textbf{(a)}). While panel \textbf{(b)} shows the classical plot of gas temperature vs. gas pressure used to characterise stellar atmospheres, the middle and lower panels illustrate the radial structures of gas temperatures \textbf{(c)}, gas densities \textbf{(d)}, gas velocities \textbf{(e)}, and condensation degrees of the element carbon into amorphous carbon dust grains \textbf{(f)}. }
\label{f:structure}
\end{figure*}

%********************************************************************
\subsection{Dynamic model atmospheres}
\label{s:DMAdetails}

For the synthetic photometry presented in this work we used the dynamic model atmospheres presented in H\"ofner et al. (\cite{HoGAJ03}). These models simulate pulsation-enhanced dust-driven winds (caused by radiation pressure on amorphous carbon dust particles), which is the most widely accepted scenario for mass loss on the AGB in the C-rich case. They are well suited to describe the complex behaviour of all atmospheric layers -- from the inner photosphere out to the cool wind region -- of pulsating AGB variables with intermediate to high mass-loss rates.\footnote{The mechanism for mass loss rates below $\approx$\,10$^{-7}$\,$M_{\odot}$\,yr$^{-1}$ (SRVs), where the $\dot M$-$P$-relation (e.g., Fig.\,1 of Vassiliadis \& Wood \cite{VassW93}, Fig.\,22 of Olofsson et al. \cite{OlEGC93}, Fig.\,7.25 of Olofsson \cite{Olofs04}) no longer holds, is still dubious (Gautschy-Loidl et~al. \cite{GaHJH04}, Mattsson et al. \cite{MatWH10}).} This is accomplished by a combined and self-consistent solution of hydrodynamics, frequency-dependent radiative transfer and a detailed time-dependent treatment of dust formation and evolution (for details see H\"ofner et al. \cite{HoGAJ03}).  Additional descriptions of the dynamical models can be found in Gautschy-Loidl et~al. (\cite{GaHJH04}), Nowotny (\cite{Nowot05}), Nowotny et al. (\cite{NAHGW05}, \cite{NowHA10}), and Mattsson et al. (\cite{MatHH07}, \cite{MatWH10}). 

In a recent study, Mattsson et al. (\cite{MatWH10}) computed a grid of C-rich dynamic model atmospheres and investigated the resulting mass loss properties (e.g., mass-loss rates $\dot M$, wind terminal velocities $u_\infty$, dust condensation degrees in the outflows $f_c$) as a function of stellar and piston parameters. Here we made use of only one selected atmospheric model, the parameters of which are listed in Table\,\ref{t:dmaparameters}. This \textit{model~S} is representative of a typical C-type Mira with intermediate mass loss and was successful in the past in simulating observational results as diverse as low-resolution spectra (Gautschy-Loidl et al. \cite{GaHJH04}), line profile variations (Nowotny et al. \cite{NAHGW05}, \cite{NoLHH05}, \cite{NowHA10}) and interferometric properties (Paladini et al. \cite{PAHNS09}).

In Fig.\,\ref{f:massenschalenS} the motions of layers at different atmospheric depths for the chosen model are illustrated, while the corresponding atmospheric structures are shown in Fig.\,\ref{f:structure}. The dynamic calculation starts with a hydrostatic initial model characterised by stellar parameters as given in the first part of Table\,\ref{t:dmaparameters}. This comparably compact and dust-free atmosphere is constructed in a similar way as classical hydrostatic model atmospheres for red giants, especially the COMARCS models used in Paper\,I (for a quantitative comparison see App.\,\ref{s:MARCSinitial}). The effects of stellar pulsation are subsequently introduced by a variable inner boundary, representing a sinusoidally moving piston at the innermost radial point of the model with parameters as listed also in Table\,\ref{t:dmaparameters}. Apart from the kinetic energy input, this also prescribes a changing luminosity input at the inner boundary (Fig.\,\ref{f:structure}a), for a detailed description we refer to Eqs.\,1 and 2 in NHA10 as well as the corresponding explanations there. A shock wave triggered by the pulsation emerges during every pulsation cycle and propagates outwards through the atmosphere causing a levitation of the outer layers. In the wake of the shocks (post-shock regions) the physical conditions -- strongly enhanced densities at temperatures low enough to allow for dust condensation -- provide the necessary basis for the formation of amC grains (e.g., Sedlmayr \cite{Sedlm94}). Radiation pressure acting upon the formed dust particles results in an outwards directed acceleration. Subsequent momentum transfer between the grains and surrounding gas via direct collisions (e.g., Sandin \& H\"ofner \cite{SandH04}) leads to the development of a stellar wind. The temporally varying radial structure of the dynamic model atmosphere (Fig.\,\ref{f:structure}) deviates significantly from the hydrostatic case. Not resembling the intital model at any point in time, the atmospheric structure of the fully developed mass-losing model becomes extremely extended in comparison, with strong local variations superposed on the shallow density gradient.

%********************************************************************
\subsection{Comparison of hydrostatic initial models and COMARCS models}
\label{s:MARCSinitial}

\begin{figure}
\resizebox{\hsize}{!}{\includegraphics[angle=90,clip]{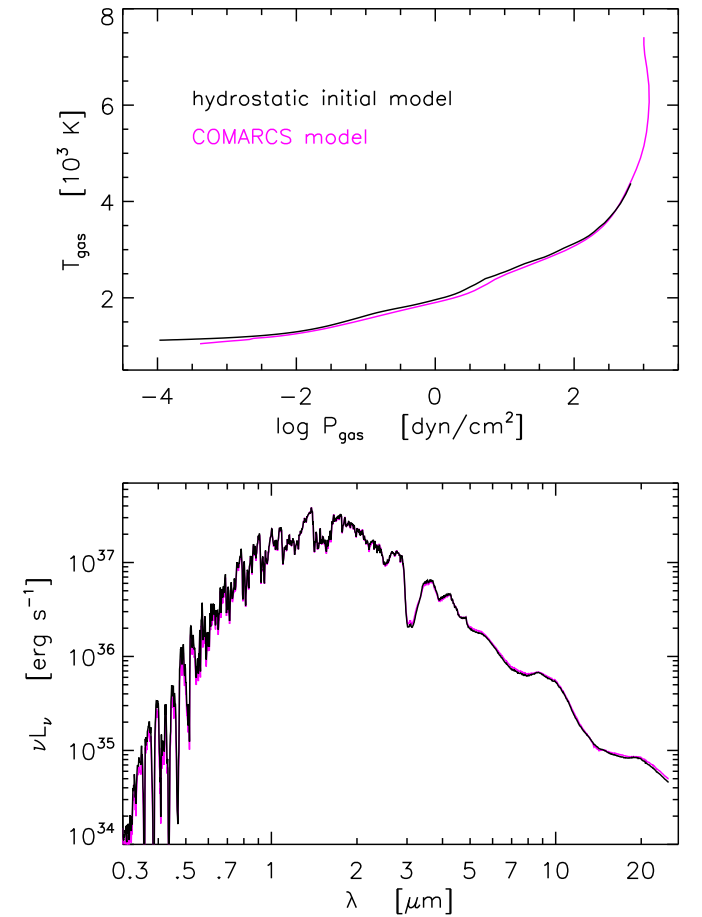}}
\caption{Comparison of the atmospheric structures of the models in Table\,\ref{t:parametercomparison} (upper panel) and the resulting low-resolution spectra based on these (lower panel).}
\label{f:VglinitMARCS}
\end{figure}

In the course of our synthetic photometry we tested how similar the hydrostatic intial models of the dynamical calculations (cf. App.\,\ref{s:DMAdetails}) are compared with classical hydrostatic model atmospheres like the ones used in Paper\,I. For the parameters of the initial model of model~S (Table\,\ref{t:dmaparameters}) no converging solution could be otained with the MARCS code for numerical reasons. Therefore, we turned to the slightly more moderate model~M used in Nowotny et al. (\cite{NowHA10}) for line profile modelling. The stellar parameters of the corresponding initial model are compared to the parameters of our closest COMARCS model in Table\,\ref{t:parametercomparison}. Note that there are marginal differences as the primary input parameters are not the same for the code to compute initial models and the MARCS code. The structures of the two atmospheric models are compared in Fig.\,\ref{f:VglinitMARCS}. Apart from the larger range covered by the COMARCS model, the radial structures are very similar for most of the depth points. This is also reflected in the resulting spectra which are shown in Fig.\,\ref{f:VglinitMARCS}, too. To quantify the differences for the different types of hydrostatic models we also compared the synthetic photometry in the chosen set of broad-band filters (Sect.\,\ref{s:synthphot}). The results, as listed in Table\,\ref{t:magcomparison}, show that the deviations are quite low. This is especially valid in the NIR. Thus, we conclude that the initial models -- which are constructed using fewer frequency points and no convection description compared to the COMARCS models of Paper\,I -- are comparable to other hydrostatic atmospheric models. They represent adequate starting points for the subsequent dynamical calculations, where the effects of pulsation and stellar winds lead to considerable changes in the radial structures (Fig.\,\ref{f:structure}).

\begin{table}
\begin{center}
\caption{Characteristics of the hydrostatic initial model used as the starting point for dynamical model~M (Nowotny et al. \cite{NowHA10}) and a very similar COMARCS model.}
\begin{tabular}{ll|cc}
\hline
\hline
\multicolumn{2}{l|}{Model:}& initial model & COMARCS \\
\hline
$L$&[$L_{\odot}$]&7000*&7015\\
$M$&[$M_{\odot}$]&1.5*&1.5*\\
$T_{\rm eff}$&[K]&2600*&2600*\\
$[$Fe/H$]$&[dex]&0.0&0.0\\
C/O ratio&\textit{by number}&1.4&1.4\\
$R$&[$R_{\odot}$]&412&413\\
\multicolumn{2}{l|}{log ($g_\star$ [cm\,s$^{-2}$])}&--\,0.614&--\,0.620*\\
\hline
\end{tabular}
\label{t:parametercomparison}
\end{center}
\tablefoot{Those quantities representing the primary input parameters are marked with asterisks.}
\end{table}      

\begin{table}
\begin{center}
\caption{Differences of photometric magnitudes calculated on the basis of the two atmospheric models of Table\,\ref{t:parametercomparison}.}
\begin{tabular}{cc}
\hline
\hline
Filter&($m_{\rm COMARCS}-m_{\rm initial}$)\\
&[mag]\\
\hline
 B & 0.276 \\
 V & 0.183 \\
 R & 0.118 \\
 I & 0.057 \\
\hline
 J & 0.005 \\
 H & 0.025 \\
 K & 0.012 \\
 L$^{\prime}$ & 0.042 \\
 M & --0.011 \\
\hline
\end{tabular}
\label{t:magcomparison}
\end{center}
\end{table}      

\end{appendix}

%####################################################################
\begin{appendix}

\section{Looping the colour}
\label{s:loopsense}

A certain disagreement was found in Sect.\,\ref{s:CMDCCD} between the modelling results and corresponding observational data. In Fig.\,\ref{f:JHKDMA} one recognises that the model loops counter-clockwise through the colour-colour plane, while the average variation of RU\,Vir in Fig.\,\ref{f:JHKObs} (dashed line) passes clockwise through an ellipse in this CCD. It can be shown by a simple test that such a change in the sense of the rotation can easily be introduced by a change of sign in the phase shift between lightcurves in different filters, which may itself be linked to the dust description in the modelling context, cf. Sect.\,\ref{s:tempvar}.

\begin{figure}
\resizebox{\hsize}{!}{\includegraphics[clip]{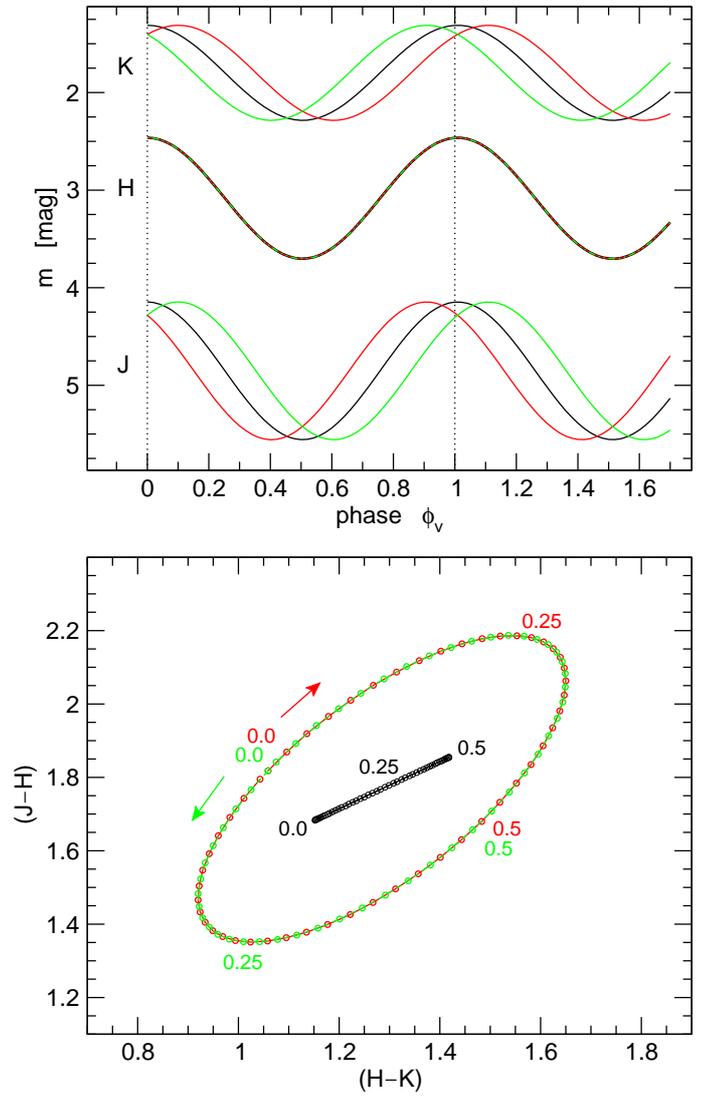}}
\caption{\textit{Upper panel:} Simulated JHK lightcurves adopted from Fig.\,\ref{f:lightcurvesRUVir} (sinusoidal fits to the observational data there) with artificial phase shifts of $\Delta \phi$~=~--0.1\,/\,0\,/\,+0.1 (red, black, and green lines, respectively) imposed in all three filters. \textit{Lower panel:} Resulting variations in an NIR colour-colour diagram, colour coded correspondingly. The labels mark locations occupied by the object at certain phases $\phi_{\rm v}$, while the arrows designate the sense of rotation of the loops.}
\label{f:loopsense}
\end{figure}

To illustrate this, we used the simulated JHK lightcurves from Fig.\,\ref{f:lightcurvesRUVir} and applied artificial phase shifts of $\Delta \phi$\,=\,$\pm$0.1 such that the phases of light maximum in the different filters occur at increasing or decreasing phases $\phi_{\rm v}$. The resulting light variations are shown in the upper panel of Fig.\,\ref{f:loopsense}.

Based on these lightcurves, we computed colour indices which are shown in the lower panel of the figure. For the JHK variations in phase (black), the object moves along a straight line from bluer to redder colours in the CCD. The ranges in ($J$\,--\,$H$) and ($H$\,--\,$K$) are determined by the amplitude differences of the individual lightcurves. If the above mentioned artificial phase shift is applied to the JHK lightcurves (red, green), the star follows a loop through the CCD which covers a larger range in both colour indices. In addition, Fig.\,\ref{f:loopsense} demonstrates that the sense of rotation changes when the sign of the phase shift $\Delta \phi$ is inverted.

\end{appendix}

\end{document}